\documentclass[lettersize,journal]{IEEEtran}
\usepackage{amsmath,amsfonts}
\usepackage{algorithmic}
\usepackage{algorithm}
\usepackage{array}
\usepackage[caption=false,font=normalsize,labelfont=sf,textfont=sf]{subfig}
\usepackage{textcomp}
\usepackage{stfloats}
\usepackage{url}
\usepackage{verbatim}
\usepackage{graphicx}
\usepackage{cite}
\usepackage{xcolor}
\usepackage{soul}
\usepackage[qm]{qcircuit}
\usepackage{textcomp}
\usepackage{soul}
\usepackage{braket}
\usepackage{url}

\begin{document}

\title{
A Primer on Security of Quantum Computing Hardware}

\author{\IEEEauthorblockN{Swaroop Ghosh$^\ast$}
\textit{The Pennsylvania State University},
PA USA,
szg212@psu.edu

\IEEEauthorblockN{Suryansh Upadhyay$^\ast$}
\textit{The Pennsylvania State University},
PA USA,
sju5079@psu.edu
\and

\IEEEauthorblockN{Abdullah Ash Saki}
\textit{The Pennsylvania State University},
PA USA,
ash.saki.e007@gmail.com

\thanks{$^\ast$Swaroop Ghosh and Suryansh Upadhyay contributed equally to this work.

$^\ast$Abdullah Ash Saki is now with IBM Quantum. The views presented in this manuscript is of his own and do not reflect the official policy or position of IBM or the IBM Quantum team.}
}





\maketitle

\begin{abstract}
Quantum computing is an emerging computing paradigm that can potentially transform several application areas by solving some of the intractable problems from classical domain. However, presence of quantum computers in the cyberspace has created security and privacy issues to this exotic computing paradigm. Similar to classical computing systems, quantum computing stack including software and hardware rely extensively on third parties many of which could be untrusted or less-trusted or unreliable. Quantum computing stack may contain sensitive Intellectual Properties (IP) that requires protection. The unique properties and requirements of quantum computing may enable launching of well-known attacks from the classical domain. For example, quantum computers suffer from crosstalk that couples two programs in a multi-tenant setting to facilitate traditionally known fault injection attacks. Similarly, frequent calibration is unique to quantum computers. Malicious calibration service providers can report incorrect error rates of qubits or mis-calibrate the qubits to degrade the computation performance for denial-of-service attacks. Quantum computers are orders of magnitude expensive than classical computers and the access queue is typically long for trusted providers due to scarce resources. Therefore, users may be enticed to explore untrusted but cheaper and readily available quantum hardware which can enable stealth of IP and tampering of quantum programs and/or computation outcomes. Recent studies have indicated the evolution of efficient but untrusted compilation services which presents risks to the IPs present in the quantum circuits. The untrusted compiler can also inject Trojans and perform tampering. Although quantum computing can involve sensitive IP and private information and can solve problems with strategic impact, its security and privacy has received inadequate attention. This paper provides comprehensive overview of the basics of quantum computing, key vulnerabilities embedded in the quantum systems and the recent attack vectors and corresponding defenses. Future research directions are also provided to build a stronger community of quantum security investigators.
\end{abstract}

\begin{IEEEkeywords}
Quantum computing; Hardware security; Security; Privacy; Fault Injection; Reverse Engineering.
\end{IEEEkeywords}

\section{Introduction}

\IEEEPARstart{Q}{uantum} Computing (QC) exploits phenomena such as superposition, entanglement, and interference to efficiently explore exponentially large state spaces and compute solutions for certain classically intractable problems. It has the potential to revolutionize numerous fields such as drug discovery\cite{cao2018potential}, chemistry~\cite{kandala2017hardware}, machine learning~\cite{qml}, and optimization. Major tech companies such as IBM~\cite{ibmqx}, Amazon~\cite{aws-braket}, and Microsoft~\cite{azure} are now offering cloud-based access to physical quantum computers. Researchers and developers are working on new algorithms and software tools to make quantum computing more accessible and usable for various industries and applications. Companies such as IBM, Pasqal, Rigetti, Xandau, and Google are developing physical hardware platforms, as well as a variety of software packages such as Qiskit\cite{wille2019ibm}, Cirq\cite{heim2020quantum}, Orquestra\cite{hevia2021quantum}, and Forest\cite{bansal2024qualitative}. 
\begin{figure}
    \centering
    \includegraphics[width=3.5in]{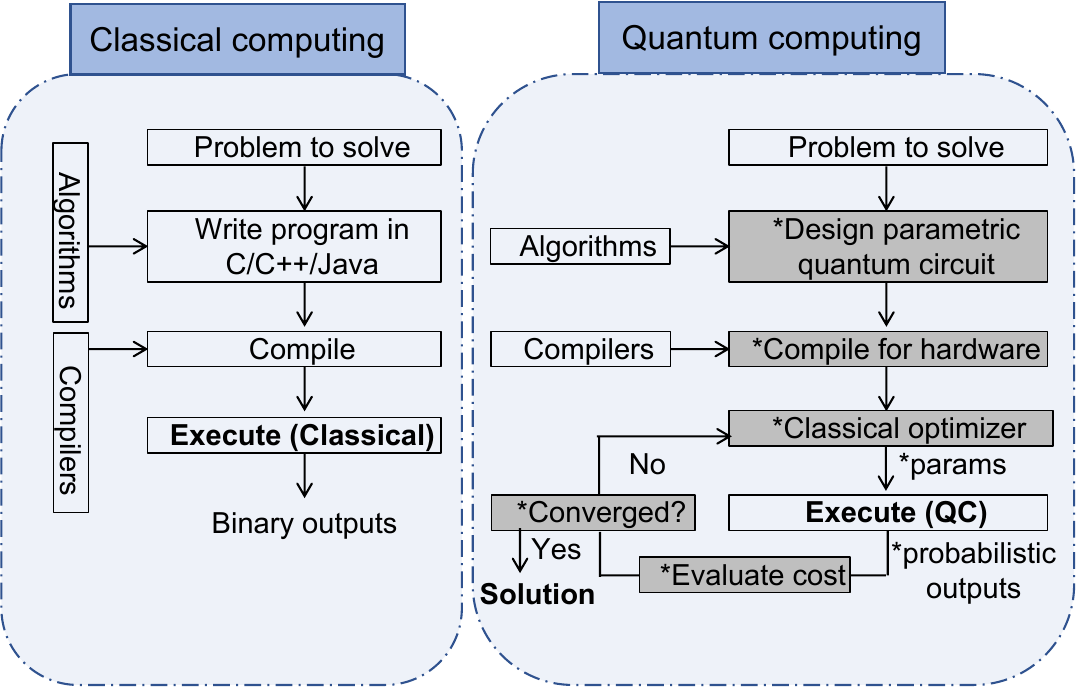}
    \caption{Comparison of classical and quantum computing paradigms, illustrating differences in computation models, data representation, and execution workflows.}
    \label{fig:clasicalvsquantum}
\end{figure}

Commercial quantum computers are already deployed in the cloud and are accessible to public. The presence of quantum computers in the cyberspace has created unprecedented security and privacy concerns. Quantum computing systems need to be secured to prevent malicious actors from exploiting their vulnerabilities, tampering with quantum algorithms/circuits, and stealing sensitive data. To raise awareness among researchers and developers about the cybersecurity of quantum computers, the Pittsburgh Quantum Institute recently hosted a workshop \cite{Workshop} supported by the National Science Foundation and the White House Office of Science and Technology Policy.

\textbf{Classical vs quantum computing:}
Classical computing works on the bits and binary/hexadecimal notation of numbers. For programming a classical computer, one follows an algorithm or heuristic to write a program to solve the problem at hand e.g., a C/C++ program using bubblesort algorithm \cite{121} to sort a list of numbers. The program is compiled to generate machine-level instructions which are processed by the computer and the output (hexadecimal or binary or character string) is returned. The inputs can either be hard coded in the program, read from a file or fed from input devices (e.g., keyboard). The output can be written to a file or sent to the output devices (e.g., display).

\begin{figure*}
    \centering
    \includegraphics[width=7in]{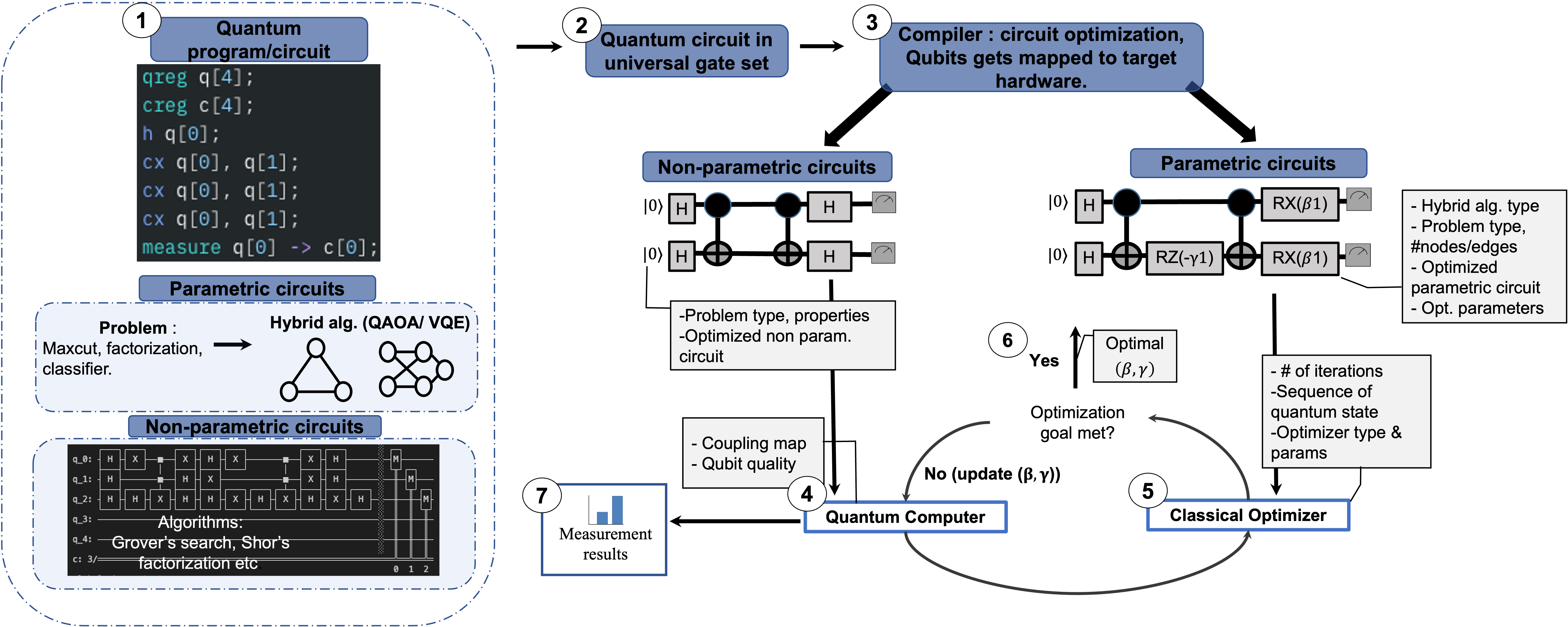}
    \caption{
    Illustration of the quantum computing workflow, from input encoding and compilation to execution on hardware. Key processes such as gate mapping, qubit allocation, and noise optimization are highlighted. Unique quantum components are marked with asterisks, and new steps are shaded in grey.}
    \label{fig:assets}
\end{figure*}

In contrast, the QC works on qubits and their entanglement, superposition and interference. Specialized quantum algorithms can solve certain class of problems whereas generic quantum algorithms can solve wide range of problems. A program is written to create a quantum circuit by following the algorithm to solve the optimization problem. The quantum circuit consists of various high-level quantum gates. For example, the variables of the problem can be converted to Ising/spin glass model \cite{122} to create the problem Hamiltonian which in turn, can be represented as a quantum circuit. Although the program could be non-parametric (i.e., without any tunable gates), it could be sensitive to quantum noise. Hybrid quantum-classical approaches have been proposed where parameters are added in the quantum circuit for iterative optimization by a classical machine (Fig. \ref{fig:clasicalvsquantum}). The inputs to the program are either encoded in the program or could be read from quantum RAM \cite{123} (if supported). The quantum programs can be written at the netlist level (using universal gates) similar to assembly language or at a higher level e.g., Q\# (similar to C) \cite{121}. The program is compiled to, (i) map universal gates to native gate set of the target hardware; (ii) ensure that the logical qubits map to physical qubits in the best possible way to minimize extra swap gates to meet the coupling constraint of the hardware; (iii) optimize the number of two qubit gates and circuit depth for resilience to quantum noise. A classical optimizer randomly initializes the parameters of the parameterized circuit before sending it to the quantum hardware. The program is executed for a certain number of times (called shots/trials) to estimate the probability of basis states. The classical optimizer computes the cost and drives the new set of parameters to optimize the objective function. After convergence, the quantum circuit with optimized parameters is expected to provide the solution to the target problem. The unique components in quantum computing are marked with asterix and the new steps are shaded in grey in Fig. \ref{fig:clasicalvsquantum}. The process of solving a problem using QC is shown in Fig. \ref{fig:assets} with few commonly used terms (e.g., coupling map). 

\begin{figure}
    \centering
    \includegraphics[width=3.5in]{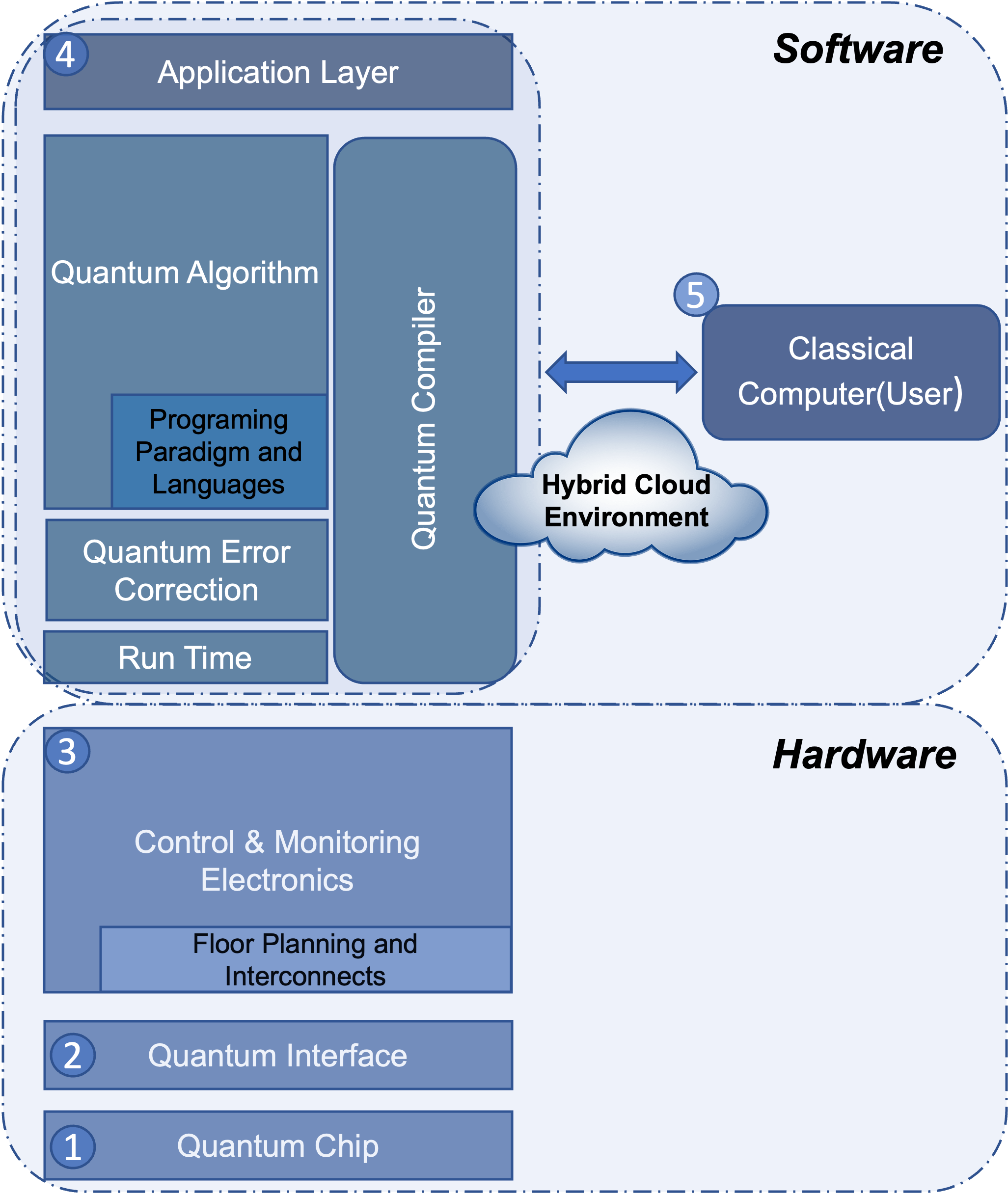}
    \caption{A comprehensive overview of the quantum stack and its components, including quantum hardware, quantum software, and hybrid cloud environment. }
    \label{fig:stack}
\end{figure}

\textbf{Current state of quantum computing:} The computing power of quantum computers is growing due to rapidly evolving noise mitigation techniques \cite{gottesman2010introduction,larose2022mitiq,shaw2021classical}, ever-increasing number of qubits and improving error rates and decoherence times. Powerful gate-based universal quantum computers can solve societal and science problems that are deemed impossible by classical computers as demonstrated by Google \cite{arute2019quantum}. IBM introduced its 133-qubit `Quantum Heron' processor, featuring an innovative architecture that substantially reduces error rates. This new processor improves upon its predecessor, the 127-qubit `Quantum Eagle'. In a separate breakthrough, Atom Computing became the first company to reach the 1000-qubit milestone in a gate-based quantum system, marking a major advancement in the field. Quantum noise presents a barrier towards applicability of Noisy Intermediate-Scale Quantum (NISQ) computers (with errors $\sim 10^{-3}$) specifically for well-known quantum algorithms that need low error rates. Fault tolerant quantum computers using quantum error correction codes (e.g., surface code) comes with hefty overhead e.g., 1000’s of physical qubits per logical qubit. Therefore, the quantum community is exploring various hybrid classical-quantum computing using shallow depth variational algorithms e.g., Quantum Approximate Optimization Algorithm (QAOA) and Variational Quantum Eigensolver (VQE) \cite{alam-cicc, alam-iccad-invited-qaoa, alam-micro, mahabubul-dac-qaoa, alam-date} to compute approximate solutions in presence of quantum noise. These hybrid algorithms employ a classical computer to drive the quantum computer iteratively to reach a solution to a given problem. The hybrid algorithms can embed security sensitive information as explained next. 

\begin{figure*}
    \centering
    \includegraphics[width=7in]{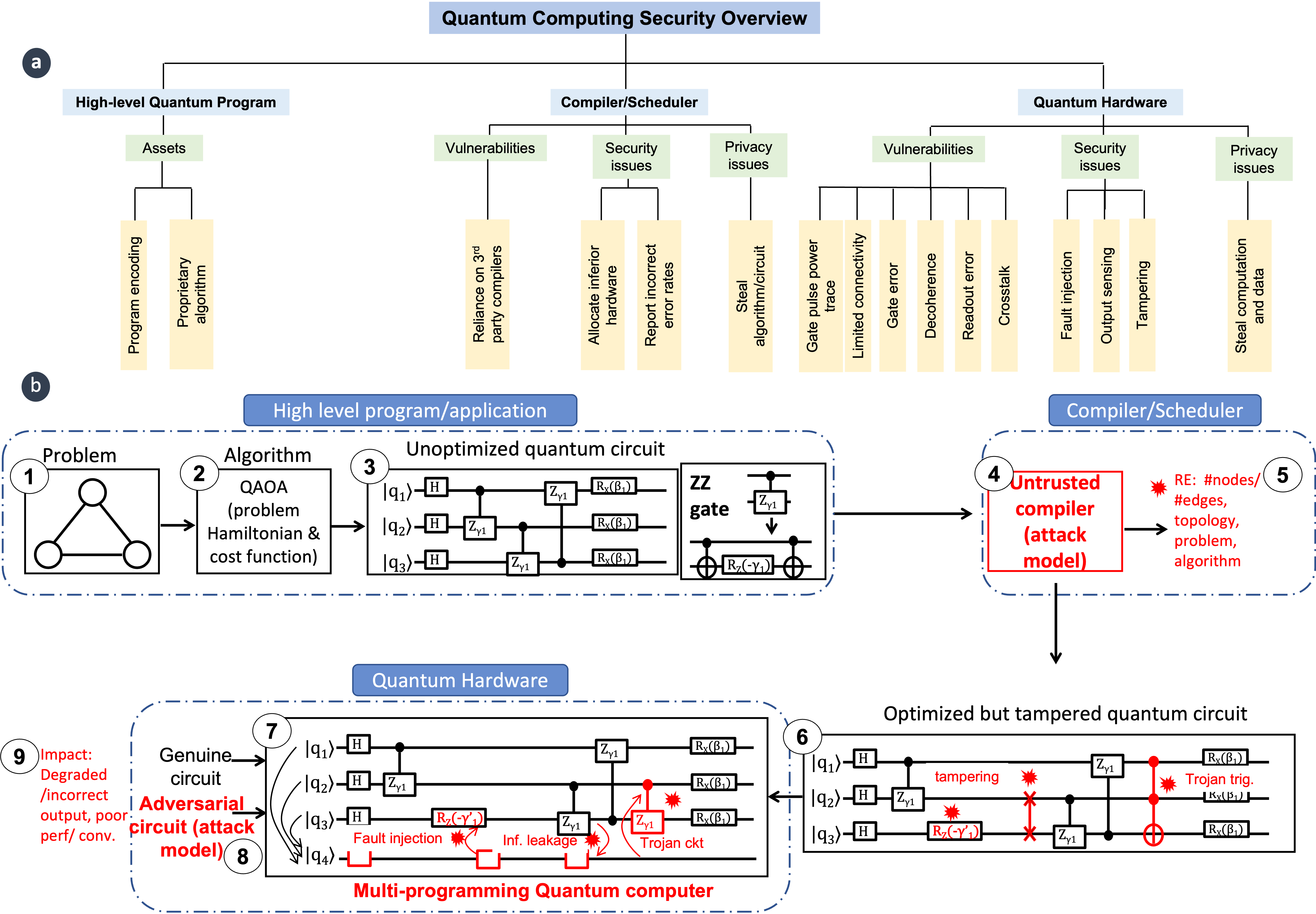}
    \caption{(a) Security taxonomy of quantum computing with some exemplary assets, vulnerabilities and security/privacy concerns. (b) Attack via untrusted compiler.}
    \label{fig:overview}
\end{figure*}

\textbf{Need for security of quantum computing:} The emerging trends in quantum computing present a host of novel security challenges for NISQ computers. For example, quantum circuits can be lucrative targets for, (i) stealth to make profit if they are reused in multiple applications e.g., Quantum Machine Learning (QML) circuits like classifiers. Problem-specific circuits optimized to solve the problem at scale can be considered as Intellectual Property (IP). Trained QML models are also orders of magnitude expensive\cite{kundu2024security} than classical machine learning (ML) models making them extremely valuable; (ii) launching various security attacks that can have strategic impact e.g., a classifier mis-prediction can have significant safety implications depending on application; (iii) leaking high-level (algorithm, problem) and low-level (data, metadata) information. In absence of understanding of the security assets, threat space and countermeasures, the unprotected quantum circuits could lose significant IP and present significant national security concerns due to poor/ incorrect optimization of problems on quantum computers that could be of national/global importance. Few exemplary security risks associated with NISQ computers are as follows:

\textit{(Problem-1) Problem-specific quantum circuits can reveal IP:} Problem-specific parametric quantum circuits designed using variational algorithms e.g., QAOA to solve certain problem embed the topology of the problem (an asset). For example, the entanglement between qubits in the QAOA circuit for portfolio management encodes the sensitive client information e.g., position (short, long, flat) and various financial constraints \cite{albareti2022structured}. Similarly, the quantum circuit for power grid (or other critical infrastructure) optimization encodes the number of nodes and their connectivity. The quantum neural network circuit contains the connectivity information, number of parameters and number of neurons. The problem information can be considered as IP that the client would like to keep confidential. {Fig.} \ref{fig:assets} depicts the assets embedded in quantum hardware and circuits. 

\textbf{Relation to classical computer security and quantum mechanical properties:} In classical computing, the structure of algorithms and circuits typically does not directly encode sensitive information about the problem being solved. They are designed to be problem-agnostic, with the sensitive information confined to the input data rather than the algorithm structure itself \cite{baumgarten2010preventing}. The problem-specific circuit design exploiting quantum mechanical properties is unique to quantum computing. For example, entanglement of two qubits is required to realize an edge between nodes that are mapped to those qubits in a graph maxcut problem solved using QAOA.

\textit{(Problem-2) Rise of untrusted but efficient compilers:} The success of a quantum circuit crucially depends on its optimization, more so than in classical systems. Even if a circuit is functionally identical to an optimized one, a poorly optimized quantum circuit will produce random outputs due to decoherence and noise. Several $3^{rd}$ party compilers~\cite{pytKET}\cite{Orquestra} claim efficient optimization of complex quantum circuits compared to trusted compilation services provided by vendors e.g., IBM, Rigetti, Amazon etc that may fail to converge for large-scale quantum circuits with higher packing density. Dependence on untrusted $3^{rd}$ party compilers for improved circuit depth, gate count and faster compilation time even for large, complex and dense quantum circuits poses risks to the assets. These compilers can be hosted on either the local machines by the $3^{rd}$ party or on the cloud service providers to launch, (i) Cloning/counterfeiting, where quantum circuit can be stolen or reproduced; (ii) Reverse Engineering (RE), where the sensitive aspects of the quantum circuit could be extracted e.g., circuit topology, problem and its properties e.g., the number of nodes and their connectivity for the maxcut problem of graphs; (iii) Tampering, where the quantum circuit or their mapping could be modified to influence the outcome e.g., mapping to poor quality qubits and/or tampering with the angles of the rotation gate can affect the output fidelity. In absence of post-compilation validation, such tampering can go undetected; (iv) Trojan insertion, where conditional triggers/ payloads can be added. Conventional testing-based detection do not apply to quantum circuits since the input is a unique quantum state and output is the solution to a NP hard problem (which is unknown unless the problem is solved); (v) Fault injection, where various parts of the circuit are deliberately mapped to crosstalk prone qubits for run-time injection of faults.

\textbf{Relation to classical computer security and quantum mechanical properties:} The potential security risks associated with using untrusted third-party compilers have long been recognized in classical computing. It has been demonstrated how a malicious compiler could introduce vulnerabilities that are extremely difficult to detect \cite{thompson1984reflections}. The classical attacks focus on modifying the functionality of the compiled code in ways that are hard to detect through normal testing procedures. However, the challenges posed by untrusted compilers in quantum computing are fundamentally different and stem from the unique properties of quantum systems. Unlike classical systems where compiler attacks primarily aim to alter functionality, in quantum computing, even subtle optimization differences can render a circuit useless. The vulnerability of quantum circuits stems from their quantum mechanical nature such as sensitivity to noise and decoherence, making optimal compilation not just beneficial but key to successful execution. The noise sensitivity creates tampering opportunities; altering quantum gate parameters, like rotation angles, can have profound effects on measurement outcomes due to the continuous nature of quantum operations. Unlike classical bits, quantum states cannot be perfectly copied due to the no-cloning theorem, adding a unique dimension to the challenge of circuit protection since tampering cannot be readily detected and corrected using classical techniques like triple modular redundancy. Moreover, the probabilistic nature of quantum measurements and the sheer complexity of simulating large quantum systems classically render conventional testing methods inadequate, necessitating new approaches to circuit validation and security.

\textit{(Problem-3) Evolution of multi-tenant compute (MTC) environment:} In MTC~\cite{multi-programming}, two or more programs share the same quantum hardware. This situation is very realistic since only ~50 qubits are needed to establish quantum supremacy~\cite{markov2018quantum} i.e., to exceed the computing power of classical supercomputers. Although the number of used qubits is problem-specific, it is likely that a large fraction of a 100+ qubit quantum hardware can remain unused. The vendor would prefer to improve the hardware utilization to recover the capital cost and/or to reduce the wait queue which is typically long due to scarce quantum resources. Note, one may argue that the 1000s of available qubits in future hardware can be exploited for error correction however, the current direction is to explore the computing power of NISQ computers using approximate algorithms to solve practical problems without any error correction. In MTC environment, the unwanted coupling between qubits i.e., crosstalk can leak information, allow fault injection ~\cite{saki-islped} and cause convergence slow down/failure (for hybrid algorithms) in victim's program by malicious programs (Fig.\ref{fig:overview}. This attack model is applicable to cloud-based quantum computing offered by major vendors e.g., IBM and Rigetti. Simple crosstalk mitigation techniques e.g., disabling all crosstalk prone qubits will not work since they can degrade the computing power of quantum computers to unacceptable levels.

\textbf{Relation to classical computer security and quantum mechanical properties:} Multi-tenant computing environments have long been a cornerstone of classical computing, particularly in cloud computing platforms rouitenly hosting multiple users' programs on shared hardware \cite{armbrust2010view}. These systems employ virtualization techniques to isolate workloads and prevent interference between users \cite{barham2003xen}. However, quantum MTC environments present unique challenges. The fundamental properties of quantum mechanics introduce vulnerabilities that are inherently different from those in classical systems. For example, the problem of crosstalk is rooted at sensitivity of qubits to noise and dissipation of excited states. Reduction of circuit depth requires packing of gates for parallel execution which in turn worsens the crosstalk. Furthermore, not all qubits are fully connected to each other for majority of qubit technologies such as superconducting and Trapped-Ion (TI) qubits. This is primarily due to quantum mechanical challenges like frequency collision and leakage. The limited qubit connectivity necessitates SWAP (for superconducting qubits) or shuttle (for TI qubits) operations for non-adjacent two-qubit gates, increasing gate count and error accumulation. Adversaries can exploit this weakness by strategically occupying qubits, forcing victim programs to suboptimal allocations with higher SWAP overhead and error rates. Crosstalk mitigation is exacerbated by the scarcity of high-fidelity qubits. Disabling affected qubits may incur significant performance penalties due to the exponential scaling of quantum computational power. These factors create a critical trade-off between maximizing gate fidelity and computational resource maximization in multi-tenant quantum environments, posing unique security challenges compared to classical systems.

\textit{(Problem-4) Rise of untrusted/less-trusted/unreliable quantum hardware:} Convergence of hybrid quantum classical algorithms such as, QAOA and training of QML algorithms rely on fast and affordable access to quantum hardware. On one hand, domestic (and likely trusted) vendors such as IBM and Rigetti are expensive and are often associated with long wait queue while on the other hand, companies such as, Baidu \cite{biadu}, the Chinese internet giant, have recently announced all-platform quantum hardware-software integration solutions, such as Liang Xi, that provide access to various quantum chips via mobile app, PC, and cloud, and connect to other third-party quantum computers. These trends lead to reliance on external (and likely less-trusted) hardware suites instead of internal counterparts. This can pose a significant risk to IP protection as untrusted/less-trusted hardware providers can steal sensitive IP and tamper with the computation outcome. 

\textbf{Relation to classical computer security and quantum mechanical properties:} The classical domain has long been familiar with the vulnerabilities associated with third-party cloud computing, such as hacking, data breaches, and insecure APIs \cite{hu2020overview}. As the quantum computing ecosystem grows, it's reasonable to expect similar, if not more complex, security challenges. The state-of-the-art quantum computers require hybrid quantum-classical approaches for noise mitigation (a quantum mechanical effect) which in turn, exposes the quantum IPs to not only quantum but also classical cloud service providers handling pre-processing of data, state preparation circuit design, compilation, post-processing of measurement outcomes and computation of objective function \cite{kundu2024security}. Furthermore, lack of encryption techniques on quantum circuits and outcomes puts the IPs and valuable results at the mercy of quantum cloud service providers. Techniques like quantum homomorphic encryption are impractical in near term due to high overhead whereas classical channels to decrypt encrypted quantum circuit is non-existent in current quantum infrastructure. Other unique aspects of quantum computing introduce new risks e.g., (a) the inability to verify results after adversarial tampering, as the correct output of a quantum program cannot be computed on a classical computer, (b) the probabilistic nature of quantum computation results, which can be manipulated by altering the basis state probabilities, and (c) low-overhead attack models, such as manipulating gate error rates, which can be challenging to detect yet significantly impact the probability of the program's correct output.

\textit{(Problem-5) IPs embedded in the quantum system:} Quantum chip and peripherals may embed multitude of IPs. For example, superconducting qubits from IBM and Rigetti may even though functionally look identical (i.e., use Josephson Junction and a capacitor), they may employ different implementation (e.g., microwave frequencies) and peripherals. The native gate sets may also differ. The hardware system and peripherals can be substantially different for another qubit technology.`Usage of third party tools and services e.g., frequent calibration of microwave pulse amplitude, phase and frequency to adjust for  dynamic noise within the quantum system present risks to these IPs. 

\textbf{Relation to classical computer security and quantum mechanical properties:}
Quantum systems, unlike their classical counterparts, embody complex physical implementations of quantum mechanical principles. While the functionality of classical chips is primarily determined by logical circuit design, the intellectual property (IP) in quantum chips extends far beyond circuit layout. It encompasses precise physical parameters such as specific microwave frequencies for qubit control, unique configurations of Josephson junctions and capacitors, and customized native gate pulses. These elements are intrinsically tied to the quantum nature of the system, where even slight variations can significantly impact qubit performance and coherence times. The diversity in qubit technologies, including superconducting, TI, and photonic systems, further complicates IP protection. Each technology requires vastly different hardware systems and peripherals, adding layers of complexity to the protection of proprietary information. Moreover, the use of third-party tools and services in quantum systems poses heightened risks compared to classical systems. These tools may require deeper access to the quantum hardware's operational parameters e.g., access to dilution fridge for calibration and access to quantum control electronics for calibration of gate pulses, potentially exposing proprietary IPs to greater vulnerability. An added challenge is lack of validation infrastructure to guarantee tamper-free calibration and/or leak of IPs.
\textbf{Contributions:} In this paper, we (i) review the basics of a quantum computing; (ii) outline various challenges and security vulnerabilities in NISQ computers; 
(iii) discuss attack models, countermeasures, and security opportunities relevant to NISQ architectures; and (iv) present future directions on quantum computer security.

\begin{figure*}
    \centering
    \includegraphics[width=7in]{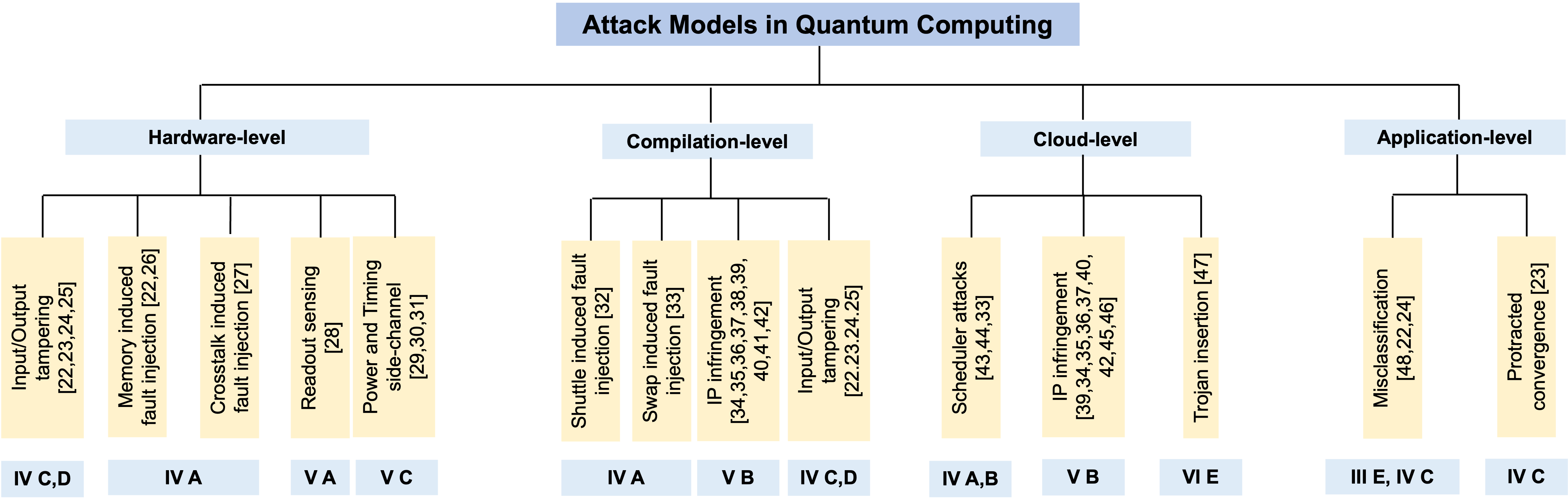}
    \caption{Quantum computing attack model taxonomy. A classification of various attack vectors targeting quantum systems, including hardware, compilation, cloud-based access, and applications.}
    \label{fig:attack_taxonomy}
\end{figure*}

In the remaining of the paper: Sections II provides background on quantum computing, Section III presents the challenges and security vulnerabilities present in the quantum computing stack. Section IV and V describe the security and privacy attack models demonstrated on NISQ devices, respectively. Section VI presents several countermeasures to thwart the attacks. Section VII draws the conclusion and presents the future outlook on security of quantum computers. An overview of the various security and privacy aspects that are reviewed in this paper is summarized in Fig. \ref{fig:attack_taxonomy}.

\section{Preliminaries}

\subsection{Qubits}

A qubit is the basic unit of quantum information in quantum computing, the quantum version of the classic binary bit physically realized with a two-state device. Electron spin, for example, can realize a qubit in which electron up-spin (down-spin) represents data '1' (data '0'). In contrast to a classical bit, which can only be either 0 or 1, a qubit can concurrently be in both $\ket{0}$ and $\ket{1}$ due to quantum \emph{superposition}. Hence, while a standard n-bit register can only represent one of the $2^n$ basis states, an n-qubit system can represent all $2^n$ basis states concurrently. A qubit state is represented as  $\varphi$ = a $\ket{0}$ + b $\ket{1}$ where a and b are complex probability amplitudes of states $\ket{0}$ and $\ket{1}$ respectively. The qubit is reduced to a single state by measurement, i.e., a pure state of $\ket{0}$ or $\ket{1}$  with probability of $|a|^2$ and $|b|^2$ respectively. Qubit states can be \emph{entangled}, allowing for the correlation of two or more qubit states. By performing a single operation on one of the entangled qubits, the states of the other entangled qubits can be changed. Furthermore, the amplitude of a qubit state can be both positive and negative. As a result, the gate operations of a quantum algorithm can be tweaked so that a negative amplitude of the same qubit state cancels out a positive amplitude (called \emph{interference}). According to researchers, quantum superposition, entanglement, and interference are at the heart of quantum speed-ups of quantum algorithms. The qubit's coefficient (or amplitude) becomes one in the read state and zero in the write state upon measurement; all information about the amplitudes is destroyed upon measurement, also known as state collapse.

\begin{figure}
    \centering
    \includegraphics[width=3.5in]{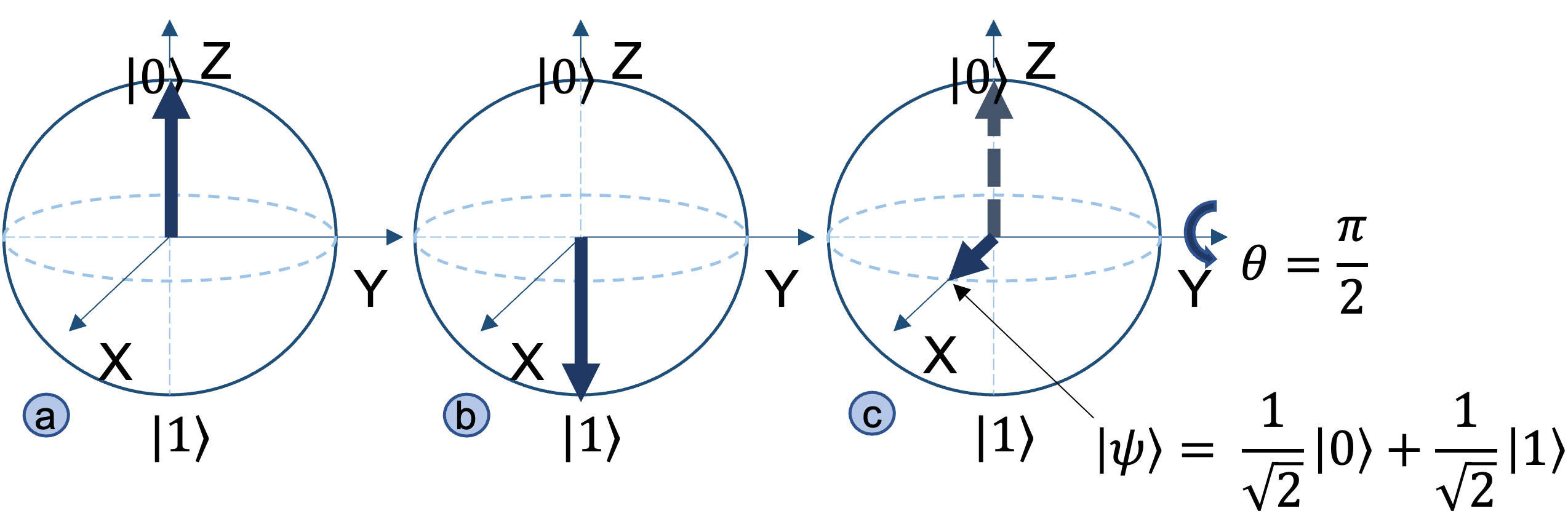}
    \caption{Bloch sphere representation of state a) $\ket{0}$ and state b) $\ket{1}$. c) Bloch sphere representation of the R$_{Y}(\pi/2)$ gate on state $\ket{0}$.  }
    \label{fig:qubits}
\end{figure}

Qubits are frequently visualized as a point on the so-called Bloch Sphere Fig. \ref{fig:qubits}. This representation shows both the phase and the probabilities of measuring a qubit as either of the basis states, which are represented as the north and south poles of the sphere.

\subsection{Quantum gates}

In QC systems, gates are utilized to manage qubit amplitudes and execute computations. Gates can act on one or more qubits at any given time. QC systems frequently support a set of universal single-qubit and two-qubit gates, analogous to classical computing's universal gates. Quantum gates, unlike classical logic gates, are implemented through the use of pulses rather than physical formation. To run a program, a sequence of gates is operated on a set of correctly initialized qubits. The gates modify the qubit amplitudes, bringing the state space closer to the desired output. Intuitively, the gate pulses cause distinct rotations along different axes in the Bloch sphere (depending on pulse amplitude, duration, and shape). Mathematically, quantum gates are represented using unitary matrices (a matrix U is unitary if UU$^\dagger$ = I, where U$^\dagger$ is the adjoint of matrix U and I is the identity matrix). For an n-qubit gate, the dimension of the unitary matrix is 2n×2n. Any unitary matrix can be a quantum gate. However, in existing systems, only a handful of gates are possible, often known as the native gates or basis gates of that quantum processor. For IBM systems, the basis gates are ID, RZ, SX, X, and CNOT. CNOT is the only 2-qubit gate, and others are single-qubit. Any non-native gate in a quantum circuit is first decomposed using the native gates.

\begin{figure*}
    \centering
    \includegraphics[width=7in]{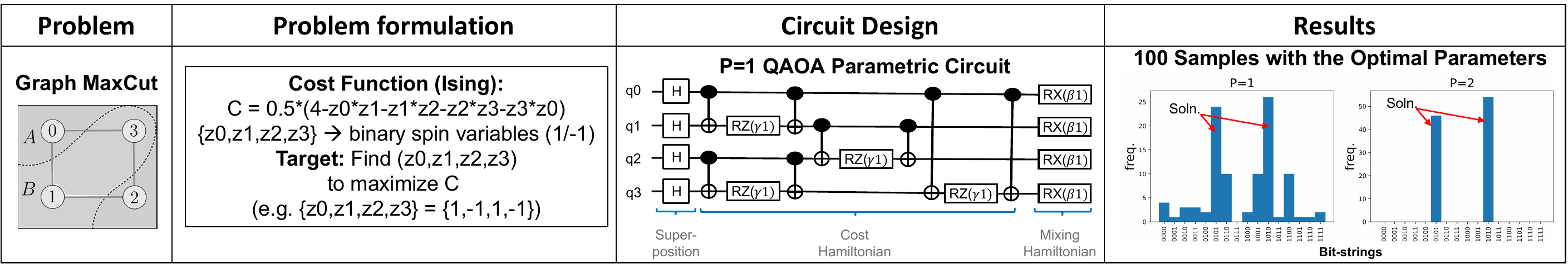}
    \caption{Example of solving a 4-node maxcut problem using quantum computing. The objective is to split the graph in two parts (1 or -1 bins) so that maximum number of edges are cut. Each node value (i.e., $z_0-z_3$) can be either 1 or -1 so we need to maximize cost function $0.5(4-z_0z_1-z_1z_2-z_2z_3-z_3z_4)$. A QAOA circuit is created to map this cost to a corresponding circuit. Each CNOT-rotation-CNOT gate combination corresponds to a cost term e.g., z0z1. The rotation parameters are optimized iteratively till the cost is maximized. The basis state providing highest amplitude (i.e., 1010) is the solution to the problem which is partition ($z_0, z_2$) and ($z_1, z_3$).
    }
    \label{fig:problem-encoding}
\end{figure*}

\subsection{Quantum circuits}

\subsubsection{Arithmetic circuits}

Many attempts have been made to build efficient classical circuits for elementary arithmetic operations ~\cite{van1990algorithms}. When the number of qubits in a quantum circuit is unimportant, it is simple to convert an efficient classical circuit into an efficient quantum circuit. A depth-efficient or size-efficient classical circuit can be converted to a depth-efficient or size-efficient quantum circuit by replacing a classical operation in the classical circuit with a corresponding quantum (in fact, classical reversible) operation. 

\subsubsection{Quantum algorithms}

Current quantum devices, limited by finite qubits and various noise sources (decoherence, gate errors, measurement errors, crosstalk), cannot fully execute quantum algorithms requiring extensive error correction, like Shor's factorization or Grover's search. However, algorithms such as QAOA and the Variational Quantum Eigensolver (VQE) show promise for near-term quantum advantage \cite{26}. VQE \cite{28}, iteratively improves approximations of a quantum chemical system's ground state energy. Starting with a tentative estimate, it computes progressively better approximations, breaking the problem into manageable components. The process repeats until meeting a heuristic stopping criterion, typically an energy threshold. This approach allows VQE to tackle complex quantum problems on current, noisy quantum devices, circumventing the need for extensive error correction and making it a viable path to quantum advantage in the near term.

Combinatorial optimization problems are addressed with QAOA ~\cite{29}~\cite{30} that uses classical optimization of quantum operations to optimize an objective function. The algorithm begins with a sequence of setup and measurement trials before being optimized by a conventional computer, similar to the VQE algorithm. The resulting quantum state after sampling provides close or exact solutions to the computation. 

\subsubsection{Quantum machine learning (QML) circuit}

Quantum Neural Network (QNN), which is the workhorse of QML models, constitutes a data encoding circuit, a parameterized quantum circuit (PQC), and measurement operations that can be trained to perform traditional Machine Learning (ML) tasks like classification, regression, and distribution generation. Numerous encoding methods are available in the literature ~\cite{schuld2021effect} among which angle encoding is the most widely used encoding scheme where a continuous variable input classical feature is encoded as a rotation of a qubit along the desired axis. Thus, `n' qubits are required to encode `n' classical features. Using sequential rotations, we can also encode multiple continuous variables in a single qubit.
The PQC contains two parts: entangling operations and parameterized single-qubit rotations. Entanglement operations are a set of multi-qubit operations performed between all of the qubits to generate correlated states. A classical optimizer iteratively optimizes the parameters to achieve the desired input-output relationship. A PQC is used by a quantum processor to prepare a quantum state. A classical optimizer is then fed an output distribution generated by repeatedly measuring the quantum state. Based on the output distribution, the classical computer generates a new set of optimised parameters for the PQC, which is then fed back to the quantum computer. The entire procedure continues in a closed loop until a traditional optimization target is met. This combination of entangling and single-qubit rotation operations is known as a parametric layer in QNN. To evaluate the efficacy of various PQC options, descriptors such as, expressive power, entanglement capability, effective dimension, and so on have been proposed~\cite{sim2019expressibility}.

\subsection{Problem formulation for quantum computing}

Combinatorial problems e.g., graph maxcut involves dividing a graph into two parts so that maximum number of edges are cut. Such problems can be formulated using spin glass/Ising model to convert the objective function to $C(z)= zTQIsingz$ for optimizing $C(z): (+1, -1)n \mapsto R_{(\geq0)}$ \cite{121}. The quantum hardware optimizes string $z*=(z1*, z2*,.., zn*)$ to maximize the cost (C(z)). An example is shown in Fig. \ref{fig:problem-encoding}. Many Quadratic Unconstrained Binary Optimization (QUBO) problems fall under this category and can be solved by following the same recipe. Another approach is to create a random ansatz (i.e., random parametric circuit) and tune the parameters to optimize the cost function corresponding to the solution. This is a generic approach and can solve any problem if the ansatz is created properly.

\subsection{Qubit technologies}

Qubits are fundamentally two-level systems, any two-level system that meets the DiVincenzo criterion \cite{nakahara2006physical} is capable of physically realizing a qubit. A number of technologies meet the qubit requirements namely superconducting, trapped ions, neutral atoms, diamond NV centers, quantum dots, and photons. Among the most common are:

\subsubsection{Superconducting qubits}

When cooled to extremely low temperatures, superconductors allow an electrical current to pass without any resistance. 
Superconducting qubits are fabricated by connecting a capacitor and a superconducting Josephson Junction (JJ) in parallel. The JJ (that acts as a non-linear inductor) requires ultra-low temperature for it to operate in the superconducting regime. Thus, superconducting qubits are usually hosted inside large dilution refrigerators. The present state of play for superconducting qubits is thoroughly summarized in Ref. ~\cite{12}. Google, Rigetti, IMEC, BBN Technologies, Intel, and IBM ~\cite{88} are notable companies pursuing research in superconducting quantum computing ~\cite{12}. 

\subsubsection{Trapped-Ion (TI) qubits}

Energy levels of electrons in neutral atoms or ions can also be used to realize a qubit. In their natural state, these electrons occupy the lowest possible energy levels. Lasers can “excite” them to a higher energy level and can assign the qubit values based on their energy state. 
Trapped ion QC system are implemented by trapping ionized atoms like Yb or Ca between electrodes using electromagnetic field ~\cite{13}. Data $\ket{0}$ and $\ket{1}$ are encoded as internal states such as hyper-fine or Zeeman states of the ions. Qubits are stored in stable electronic states of each ion, and quantum information can be transferred through the collective quantized motion of the ions in a shared trap (interacting through the Coulomb force). IonQ, Honeywell, Alpine Quantum Technologies, and Universal Quantum are notable companies pursuing research on trapped ion quantum computing. TI systems typically employ a single trap design, which has significant scaling issues. A modular design known as the Quantum Charge Coupled Device (QCCD) has been proposed \cite{39} to advance toward the next significant milestone of 50–100 qubit TI devices.

\subsubsection{Spin qubits}

Controlling the spin of charge carriers (electrons and holes) in semiconductor devices can also be used to implement a Qubit \cite{14}. Most quantum particles behave like little magnets. This trait is known as spin, the spin orientation is always either fully up or fully down. A spin qubit is created by combining these two states. Local depletion of two-dimensional electron vapors in semiconductors such as gallium arsenide, silicon, and germanium has been used to create spin qubits. Some reports also show implementation in graphene \cite{15}.

\subsubsection{Neutral Atom qubits}

Neutral atom quantum computing systems utilize individual neutral atoms as qubits, typically trapped in arrays using optical tweezers or optical lattices created by laser beams \cite{henriet2020quantum}. The qubit states $\ket{0}$ and $\ket{1}$ are usually encoded in hyperfine ground states of the atoms. Unlike trapped ions, neutral atoms interact weakly with each other in their ground states, which can be advantageous for reducing unwanted interactions. However, strong interactions can be induced on-demand by exciting the atoms to Rydberg states - highly excited electronic states with large dipole moments. Companies such as Atom Computing, QuEra Computing, and ColdQuanta are at the forefront of developing neutral atom quantum computers.

\section{Vulnerabilities in quantum computing}

This section covers quantum computing vulnerabilities at the technology, device, circuit, system, and program-specific levels. These vulnerabilities represent potential entry points for adversarial exploitation, independent of specific attack methodologies.

\subsection{Vulnerabilities at the technology level}

The security of qubits and quantum systems is a complex and ongoing area of research. Some of the security risks associated with common qubit technologies include:

\subsubsection{Limited qubit lifetime and errors}

Quantum computing relies on the stability of qubits to perform computations. However, the fragile nature of qubits makes them susceptible to errors, which can impact their performance and reliability. The lifetime of a qubit is directly related to its coherence time, which is the length of time that the quantum state of a qubit remains coherent and stable. In superconducting qubits, for example, the coherence time is influenced by the quality of the superconducting material, the temperature of the environment, and the presence of any magnetic fields. In trapped ion qubits, the coherence time is influenced by the trapping potential, the laser power and stability, and the presence of any stray electric or magnetic fields. The storage of information in qubits can be disrupted by spontaneous losses called \emph{decoherence}. This occurs when a qubit, for instance, in state $\ket{1}$, interacts with its surroundings and loses energy, leading to a state $\ket{0}$. This process is referred to as \emph{relaxation}. Another form of qubit state loss is \emph{dephasing}, where a qubit loses its phase information. Both relaxation and dephasing can be quantified by T1 and T2 times, respectively.

Quantum gates and measurement operations on qubits can also be erroneous. These are referred to as \emph{gate error} and \emph{measurement/readout error}, respectively. An incorrect logical output of a quantum gate may result from readout error, while measurement error can flip a qubit state, e.g. recording a 0 as a 1 or vice versa. The severity of gate and readout error can be measured by the gate error rate and measurement error rate, respectively. 
Crosstalk is another source of error in quantum devices, causing increased gate error in parallel gate operations. It is important to note that all forms of errors, including decoherence, gate error, measurement error, and crosstalk, are subject to temporal variation. Adversaries can launch attacks by exploiting noise sources like crosstalk and/or readout error that correlate qubit operations.

\begin{figure}
    \centering
    \includegraphics[width=3.5in]{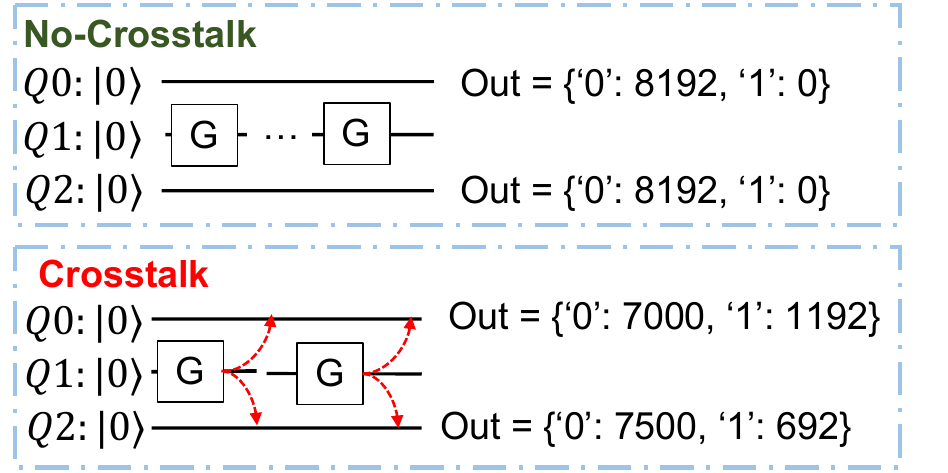}
    \caption{The impact of crosstalk on neighboring qubits during gate operations on qubit: Q1.}
    \label{fig:crosstalk}
\end{figure}

\subsection{Vulnerabilities at the device level}

\subsubsection{Crosstalk}

Crosstalk refers to the unintended transfer of quantum information between two or more qubits in a quantum circuit. It occurs when the interactions between the qubits interfere with each other, causing information to leak from one qubit to another. Various sources of classical crosstalk in Transmon-based quantum computers, such as electromagnetic crosstalk between microwave lines and stray on-chip electromagnetic fields, have been outlined in a study \cite{sarovar2019detecting}. It can arise from various sources, such as the interaction of qubits through mutual capacitive coupling or the use of shared control lines. Magnetic flux control fields used to operate qubits have been shown to affect unaddressed qubits, causing crosstalk in superconducting qubit systems \cite{flux-line-crosstalk}. Crosstalk is a major challenge in the design and implementation of large-scale quantum circuits. It has been demonstrated \cite{saki-tqe} that existing works on error resilience prioritizing only the gate error do not adequately capture NISQ behavior, and that incorporating crosstalk in quantum circuit simulation improved accuracy. This work also established crosstalk as a common source of error in NISQ devices. Fig. \ref{fig:crosstalk} depicts one such example of crosstalk that can affect circuit output by introducing errors. The final state should be the initial state with no gates applied to qubits Q0 and Q2. However, we observe errors in the measured output states of Q0 and Q2 due to crosstalk from gate operation on qubit Q1. Crosstalk can be a significant issue in a MTC environment, where multiple quantum programs are executed simultaneously on different physical qubits. Fault-injection attacks can be launched by manipulating the interactions between qubits. The main objective of these attacks is to introduce errors into the computation.

\subsubsection{Hardware variability}
In quantum computing, hardware variability refers to variations in performance of various quantum devices, such as, differing gate error rates and decoherence times. These variations arise from various sources including fabrication processes, environmental conditions, and device-to-device variation. The hardware variability can affect the performance and stability of quantum computers, resulting in errors and fluctuations in qubit parameters that can impact the accuracy of quantum algorithms. 
To mitigate the variability, advanced error correction techniques are being investigated. These methods include encoding a logical qubit with multiple physical qubits, real-time correction of qubits, and implementing robust fabrication processes to reduce variability in the qubits. However, hardware variability creates opportunities where an adversary could allocate inferior quantum hardware to a user instead of the desired hardware. This type of attack, known as a scheduler attack, can be executed by exploiting the lack of transparency in the allocation process of quantum resources.

\subsubsection{Quantum computer power side channels}
Quantum computers are predominantly cloud-based, and users typically do not have direct control over the physical environment. Despite the trustworthiness of the cloud provider or a lack of incentives for spying on users, malicious insiders or other attackers may use side-channels to extract information about algorithms executed on these computers. While the superconducting qubits themselves are isolated in a cryogenic refrigerator, an adversary can target the control electronics instead. In superconducting qubit machines from IBM, Rigetti, or other providers, RF pulses are utilized to execute gate operations on single qubits or two-qubit pairs. These control pulses are classical and vulnerable to potential eavesdropping. 
The power consumption information from the control electronics can be used to reconstruct quantum circuits and extract sensitive information about the executed algorithms, assuming physical access to the control electronics. Power side channels can also be leveraged to to recover information about the control pulses sent to quantum computers \cite{xu2023exploration}.



\begin{figure}
    \centering
    \includegraphics[width=3.45in]{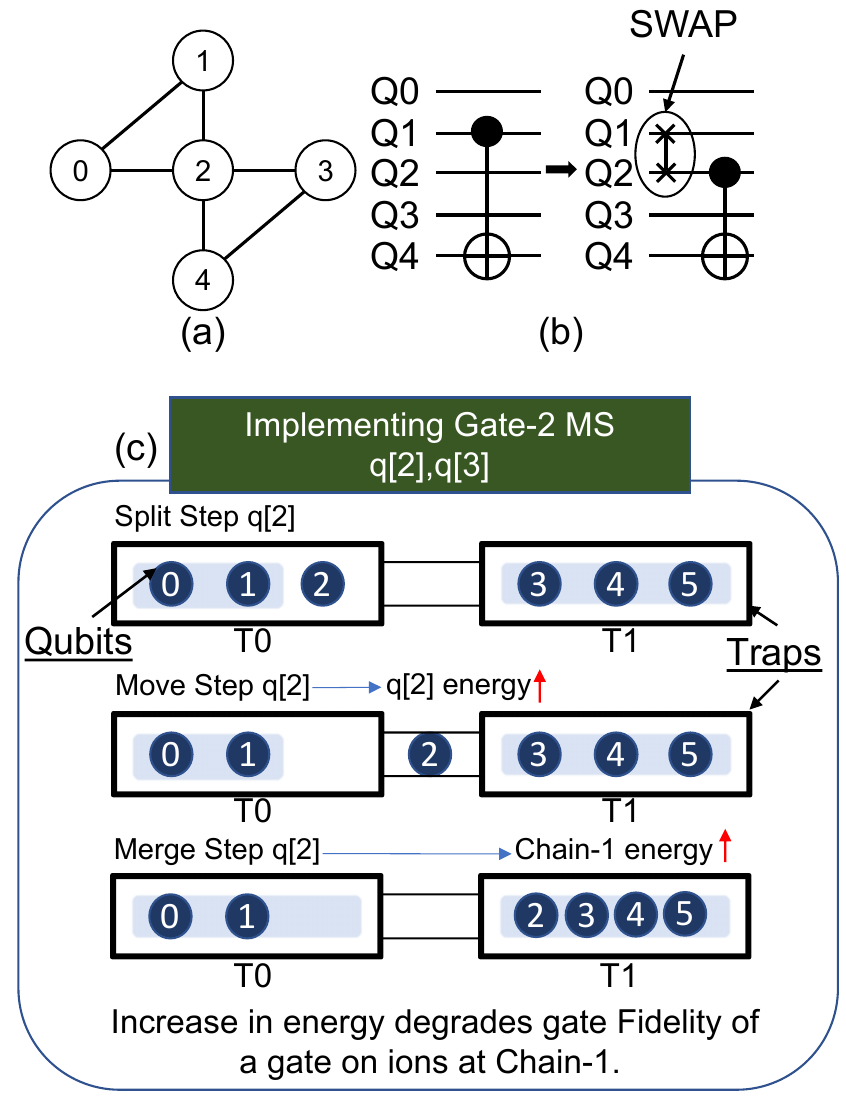}
    \caption{(a) The coupling map of ibmq\_5\_yorktown. (b) A SWAP gate is inserted to satisfy coupling constraint as Q1 and Q4 are not connected. (c) Shuttle steps to move ion-2 from trap T0 to trap T1.}
    \label{fig:swap}
\end{figure}

\subsection{Vulnerabilities at the circuit level}



\subsubsection{Coupling constraint}\label{subsec:coupling-constraint}
One of the biggest challenges in building practical quantum computers is the management of qubit interactions, also known as \emph{coupling constraint}. In superconducting systems, the qubits are implemented using superconducting circuits, and the interaction between qubits is mediated by a coupler, which typically consists of a microwave resonator. The distance between the qubits and the strength of the coupler determine the interaction between qubits. Due to the limited strength of the coupler, direct interaction between qubits is often not possible. The limited connectivity prevents 2 qubit gates between any two arbitrary qubits. For superconducting systems a compiler needs to add SWAP operations to satisfy the coupling constraint. Fig.~\ref{fig:swap}a shows the coupling graph of an IBM quantum computer where we cannot perform a CNOT (CX) gate directly between Q1 and Q4 as they are not connected. One option is to \emph{SWAP} Q1 and Q2 so that the data of Q1 moves to Q2. Now, the CX can be applied between Q2 and Q4 as they are connected. A SWAP operation includes 3 CX gates (when translated to the native instructions of the device) which increases the run-time of the quantum program and the gate count that negatively affects the circuit performance due to qubit lifetime and gate error. Adversary can exploit this vulnerability in MTC environment by occupying qubits strategically and forcing more SWAP gates in victim's program \cite{upadhyay2024stealthy}.

In trapped ion systems, qubits are implemented using ions trapped in a linear or planar trap. The physical separation of the ions between two different traps makes direct interactions difficult. A compiler adds shuttle operations to a quantum program to satisfy the inter-trap communication, however, it increases program execution time and degrades quantum gate fidelity (usually defined as the complement of the error rate). A lower gate fidelity will introduce more errors in the output and can completely decimate the result \cite{saki2021shuttle}. The shuttle operation involves several steps as shown in Fig.~\ref{fig:swap}c. For example, gate MS q[0], q[1] utilizes ions from trap T0 and can be performed without the need for additional operations. Conversely, executing the gate MS q[2], q[3] requires ions from two different traps, necessitating a shuttle operation to bring them together. During the shuttle operation, ion-2 is separated from Chain-0, transported from T0 to T1, and subsequently merged with Chain-1. The energy required for the shuttle operation can result in reduced gate fidelity, underscoring the importance of minimizing the number of shuttle operations required. In a MTC environment, security issues can arise due to qubits from an adversary's program potentially spanning over multiple traps and sharing a trap with qubits from a victim's program \cite{saki2021shuttle}. The adversary can exploit this by designing their program to require frequent shuttles between traps, which can add energy to an ion and increase the ion-chain's energy. This can result in degraded reliability of computation, characterized by reduced gate fidelity.

\subsubsection{IP risks in quantum problem encoding}

Problem encoding refers to the process of mapping a classical computational problem onto a quantum circuit. This is a crucial step in quantum computing as it determines how the problem will be solved using quantum algorithms. However, the construction of a quantum circuit can reveal sensitive information about the problem, making it vulnerable to security threats.
For example a combinatorial maxcut problem which involves dividing a graph into two parts so that the maximum number of edges is cut can be formulated using the spin-glass/Ising model. Fig.~\ref{fig:problem-encoding} shows the QAOA circuit for the problem graph in Fig.~\ref{fig:problem-encoding}. Here, each edge in the graph is represented by a CNOT-rotation-CNOT gate in the circuit. Therefore, one can infer the problem graph just by studying the circuit. Besides MaxCut, many Quadratic Unconstrained Binary Optimization (QUBO) problems fall under this category. 


\subsection{Vulnerabilities at the system level}\label{subsec:cloud}

\subsubsection{Vulnerabilities due to cloud-based access}

Quantum computing necessitates specialized resources such as cryogenic temperature~\cite{krantz2019quantum, bruzewicz2019trapped} and high vacuum all of which are currently in short supply, preventing it from becoming a personal commodity. As a result, cloud-based access to quantum computers (hosted in a remote location) is the logical next step. This allows users to interact with the quantum computers remotely, eliminating the need for physical access to these complex and expensive systems. To access cloud-based quantum computers, users typically log in to a web interface that provides a secure connection to a remote quantum computer. 
Currently, IBM, Google, Microsoft, Qutech, QC Ware and AWS Braket are some of the top vendors that provide access to quantum hardware (both superconducting and TI qubits) to the users over cloud. The reliance on high-speed, low-latency networks for remote access creates potential points of interception or disruption, raising network security risks. Additionally, sensitive information and quantum programs transmitted over the cloud are at risk of unauthorized access or IP theft, leading to data privacy concerns \cite{upadhyay2023trustworthy, saki2021split}. Users also face hardware integrity uncertainties, as they have limited ability to verify the quality and specifications of the quantum hardware they're using remotely. Scheduler vulnerabilities present another challenge, as the complex task of allocating quantum resources efficiently can be exploited, potentially leading to suboptimal qubit assignments or inaccurate error reporting \cite{upadhyay2022shuttle, puf}. Cloud-based quantum computing platforms also face challenges like job submission latency and queue backlogs, impacting hardware utilization efficiency. As quantum hardware grows in size and improves in qubit error rates, MTC has emerged as a potential solution. However, NISQ computers present unique resource-sharing challenges related to qubit connectivity, program isolation and program allocation \cite{upadhyay2024share}.

The emergence of Quantum Machine Learning as a Service (QMLaaS) represents a significant step in making quantum computing more accessible. However, this cloud-based paradigm introduces a range of vulnerabilities that span both classical and quantum domains \cite{kundu2024security}. Confidentiality is a primary concern, with risks including data theft during classical preprocessing, potential extraction of encoding circuits, and the possibility of intellectual property theft, particularly of novel quantum architectures \cite{upadhyay2024quantum,upadhyay2025quantum}. The quantum hardware itself is not immune to vulnerabilities, being susceptible to side-channel attacks and state leakage, which could potentially expose sensitive information. Integrity threats pose another significant challenge for QMLaaS. These include the risk of data poisoning attacks on training data, which could compromise the accuracy and reliability of quantum machine learning models. Adversarial obfuscation of quantum circuits and fault injection attacks exploiting qubit crosstalk are additional integrity concerns that could manipulate the behavior of quantum algorithms. Availability is also a critical issue in the QMLaaS landscape. The potential for Denial of Service (DoS) attacks targeting both classical and quantum resources could disrupt service and impede access to quantum computing capabilities \cite{mahjabin2017survey}. Ransomware attacks encrypting vital data and intentional latency injection in circuit execution are additional threats that could compromise the availability and performance of QMLaaS systems \cite{brewer2016ransomware}.

\subsubsection{Risks of using untrusted compilers}

The optimization of the circuit for enhanced circuit depth and reduced gate count is an important part of quantum circuit compilation. Several third-party compilers are emerging that offer optimization at faster compilation times even for large quantum circuits\cite{zulehner-a*, muqut}. The following considerations may prompt quantum circuit designers to utilize untrusted third-party compilation services: (a) Optimized quantum circuits: it is crucial to have well-optimized quantum circuits otherwise it may produce random outputs even if functionally correct. (b) Limited availability of trusted compilers: there may be a shortage of trusted compilers that have kept up with the latest optimization advancements. (c) 
Cost savings: utilizing an untrusted compiler may offer cost savings compared to using a trusted compiler, which could be particularly appealing for individuals or organizations with limited budgets. (d) Faster turnaround times: untrusted compiler may result in faster turnaround times potentially due to access to more resources or a better streamlined process compared to trusted compilers.


The use of third-party quantum circuit compilers, while offering potential benefits in optimization and efficiency, introduces several vulnerabilities. One primary concern is intellectual property exposure, as untrusted compilers may have access to proprietary quantum algorithms or circuit designs \cite{saki2021split}. There's also a risk that untrusted compilers might not optimize circuits as claimed, potentially impacting the performance and accuracy of quantum computations, which raises questions about optimization integrity.
Malicious compilers could introduce subtle changes to the quantum circuit that are difficult to detect. Advanced compilation techniques might inadvertently reveal sensitive aspects of the quantum algorithm's structure or purpose, leading to reverse engineering risks. Furthermore, the accuracy and consistency of results from circuits optimized by untrusted compilers may be questionable, introducing reliability concerns \cite{obfuscation-aks,saki2021split}.

\subsubsection{Risks of using untrusted hardware providers}
Consumers have the option of utilizing quantum hardware through cloud services. However, high cost, and long wait times may necessitate the use of emerging/untrusted/less-trusted/unreliable quantum hardware providers. The prices charged by AWS Braket, IBM, and Google Cloud ranges from $\$$0.35 to $\$$1.60, based on the qubit count, for a runtime of 1 millisecond per shot which is expensive. To factor a 2048 bit product of two primes, for example, a quantum computer will need approximately 25 billion operations in 14238 logical qubits \cite{gidney2021factor}, which equals 432 billion qubit-seconds. Along with high cost, quantum computers often results in lengthy wait times. For IBM quantum systems, only about 20$\%$ of total circuits have ideal queuing times of less than a minute \cite{ravi2021quantum}. The average wait time is about 60 minutes. Furthermore, more than 30$\%$ of the jobs have queuing times of more than 2 hours, and 10$\%$ of the jobs are queued for as long as a day or longer! As the quantum computing ecosystem continues to advance, an emergence of third-party service providers is likely that can potentially offer better performance. This may attract users to these services. For instance, there are third-party compilers, such as Orquestra \cite{Orquestra} and tKet \cite{pytKET}, that are now available and can support hardware from multiple vendors. Baidu's ``Liang Xi" \cite{biadu} provides flexible quantum services through private deployment, cloud services, and hardware access, and can also connect to other third-party quantum computers. These trends can result in a dependence on third-party compilers, hardware suites, and service providers that may not be as reliable or secure as some trusted alternatives. 

A primary concern is hardware integrity uncertainty, as users have limited ability to verify the actual specifications and performance of quantum hardware provided by less-established vendors. This uncertainty extends to data confidentiality risks, where utilizing untrusted hardware providers may expose sensitive quantum computations or IPs to unauthorized access or theft. Additionally, reliability of computation can also be a significant concern, as less reliable quantum hardware may produce inconsistent, tampered or inaccurate results potentially compromising the integrity of quantum computations. Service continuity vulnerabilities are another consideration, as emerging providers may not offer the same level of service stability and support as established vendors, potentially leading to interruptions in quantum computing capabilities.

\begin{figure}
    \centering
    \includegraphics[width=3.5in]{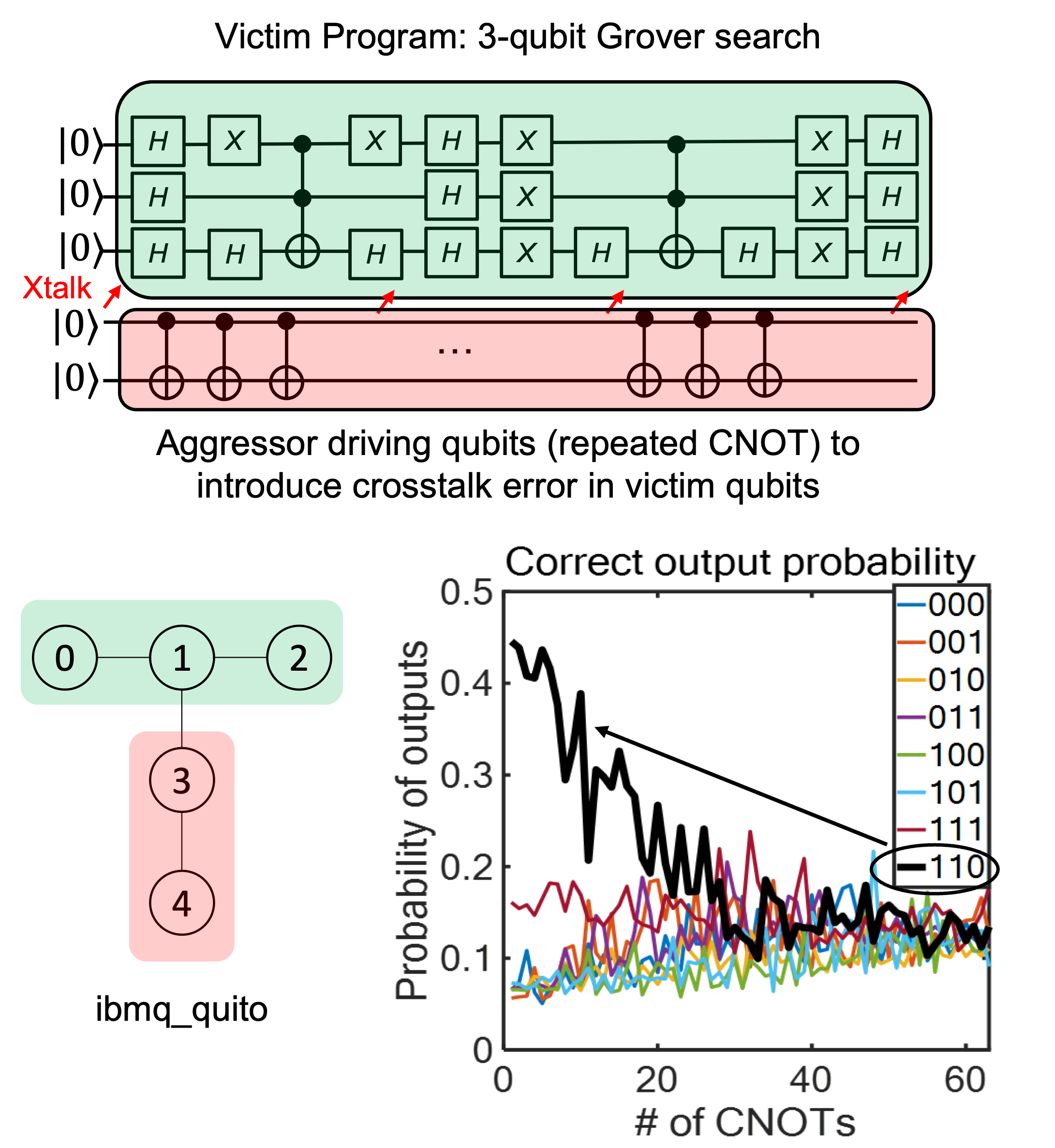}
    \caption{A 3-qubit Grover search circuit under attack with a conceptual diagram of a crosstalk-induced fault-injection attack \cite{saki-islped}.}
    \label{fig:fault-injection}
\end{figure}

\begin{figure*}
    \centering
    \includegraphics[width=7in]{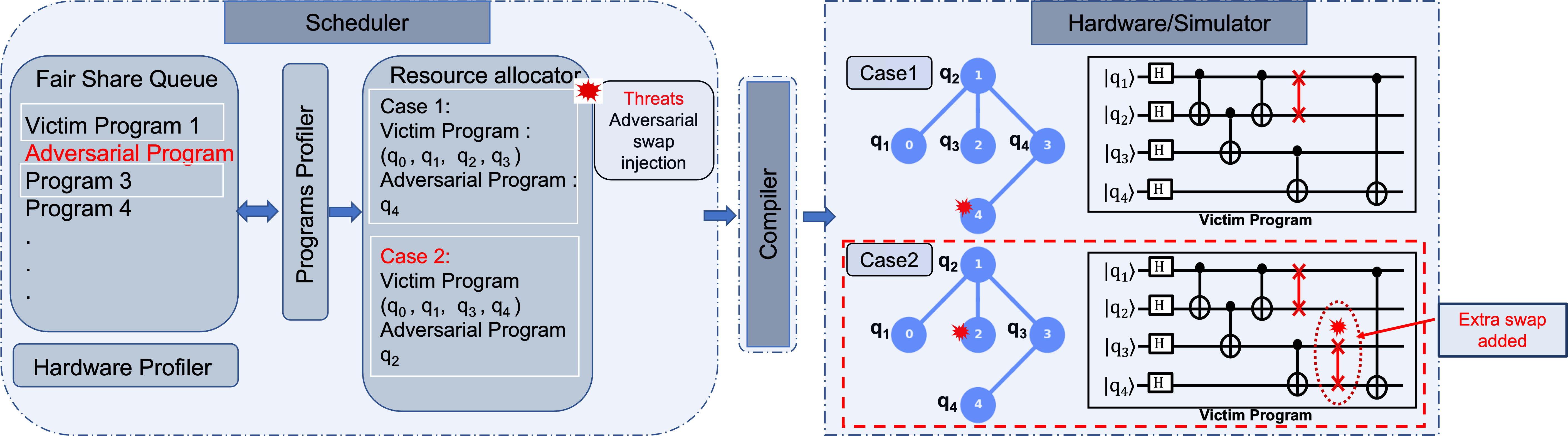}
    \caption{Threat model in MTC: The scheduler determines job execution order and concurrent program allocations on the quantum hardware for a single run in MTC. An adversary can exploit scheduler oversights and quantum hardware connectivity constraints. We compare two cases \textbf{Baseline (Case-1):} The victim program is allocated qubits 0, 1, 2, and 3, while the adversary program is allocated qubit 4. \textbf{Adversarial Attack (Case-2):} The victim program is allocated qubits 0, 1, 3, and 4, while the adversary program strategically occupies qubit 2. The attack involves extra SWAP gates introduced on the victim circuit.
    }
    \label{fig:swap-injection}
\end{figure*}

\subsection{Program agnostic vulnerabilities}

\subsubsection{Vulnerabilities in hybrid algorithms}

Variational algorithms like QAOA require the design of problem-specific parametric quantum circuits to solve particular problems, making the circuit's topology a valuable intellectual property (IP) or asset. This is especially true in sensitive applications such as power grid or critical infrastructure optimization, where clients may need to maintain confidentiality about problem information. Hybrid quantum-classical algorithms, including QAOA and VQE, are characterized by their unique workflow involving input parameters, quantum circuit execution, measurement and readout, classical post-processing, and iterative convergence. While powerful, these features also introduce potential vulnerabilities. As third-party service providers offering higher performance quantum computers become more prevalent, and NISQ computers continue to scale, the risk of IP infringement and other security concerns increases. A critical vulnerability in these algorithms stems from the potential for malicious tampering at various stages of the hybrid workflow. Subpar hardware or intentional tampering by the hardware provider can prolong convergence times, potentially leading to increased computational costs or degraded solution quality \cite{upadhyay2023trustworthy}. Fault injection in the measurement process or parameter tampering can compromise the integrity of the results, making the algorithm less efficient or steering it towards sub-optimal solutions \cite{upadhyay2022robust}. Currently, there are no effective ways to detect or prevent tampering with input parameters or output results in hybrid quantum-classical algorithms.

\subsubsection{Vulnerabilities in QML}

Quantum machine learning (QML) is an emerging field that aims to develop quantum algorithms to perform conventional generative/discriminative machine learning tasks (e.g., classification, regression, etc.) \cite{qml, killoran2019continuous, dallaire2018quantum, schuld2020circuit}. However, the security of QML systems is still an area of active research and development, and there are several potential vulnerabilities that need to be addressed. One significant challenge is the presence of $barrenplateaus$ in the training landscape of parameterized quantum circuits \cite{mcclean2018barren}. These areas of vanishing gradients can severely hinder the training process, causing gradient-based optimization methods to stall and preventing effective model learning. The impact of quantum noise on QML systems presents another vulnerability. It has been noted \cite{wang2020noise} that quantum noise can induce $barrenplateaus$ in the training landscape, further complicating the training process. Additionally, temporal variations in quantum noise can affect the reliability of quantum classifiers, potentially leading to inconsistent or incorrect classifications over time \cite{mahabubul-islped}. QML systems have also demonstrated vulnerability to adversarial attacks \cite{liu2020vulnerability}, where minimal alterations to input data can result in misclassification. This vulnerability becomes more pronounced in high-dimensional quantum classifiers, where even small perturbations can lead to significant errors.

\begin{figure}
    \centering
    \includegraphics[width=3.5in]{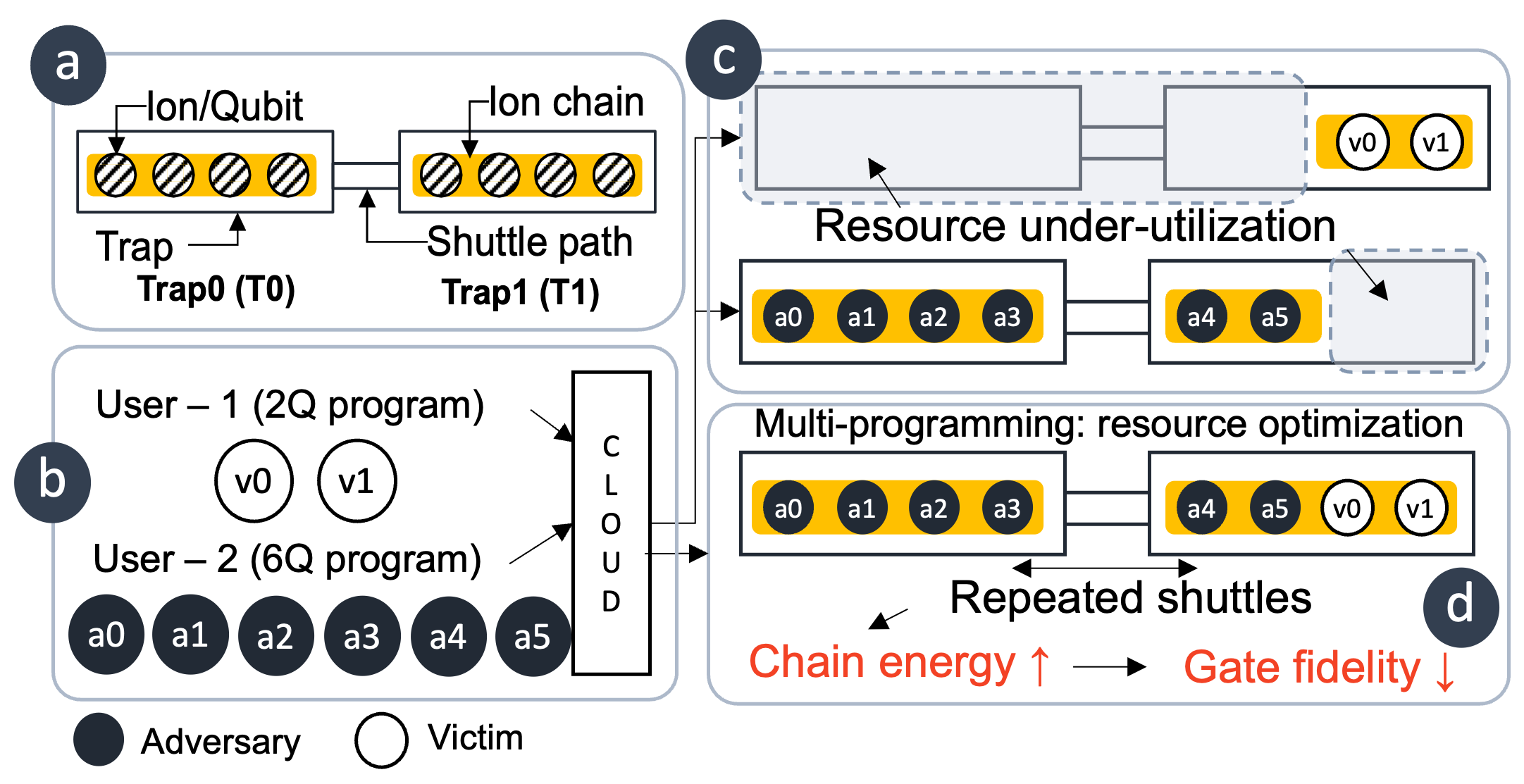}
    \caption{ (a) Overview of a TI system. (b) The basic concept of
multi-programming. (c) Issues with single programming. (d) Attack
overview: the malicious program utilizes a shared trap with the victim program. The attacker aims to induce repeated shuttles between traps, resulting in elevated chain-energy and a decline in gate fidelity within the shared trap \cite{saki2021shuttle}.}
    \label{fig:TI}
\end{figure}

\section{Security issues in quantum computing}

This section presents various security threats to the quantum computing ecosystem such as fault injection, scheduler attacks, input/output tampering, and threats to privacy Fig. \ref{fig:attack_taxonomy}.

\subsection{Fault injection}

\subsubsection{Fault injection via control electronics
}

Vulnerabilities in FPGA memories that are used to store amplitude and phase information for generating quantum gate pulses have been explored for fault injection \cite{das2024investigating}. It has been noted that bit flip errors in FPGA memories, due to electromagnetic interference, power fluctuations, and temperature variations, as well as potential adversarial fault injections, can lead to errors in quantum gate operations. It has been shown that bit flips in the exponent and initial mantissa bits of the real amplitude cause substantial deviations in quantum gate operations, with Total Variation Distance (TVD) increases as high as $\approx200\%$.

The attack model and the explored vulnerabilities are unique to quantum computing due to the analog nature of quantum control signals and the extreme sensitivity of qubits to small variations in these signals. Unlike classical digital systems where small errors might be tolerated or corrected by discretization, even minor alterations in quantum control pulses can lead to significant errors in quantum operations. The need for precise analog control signals in quantum computing introduces a new attack surface which was absent in classical digital systems. In classical systems, bit-flip attacks on memory have been studied \cite{kwong2020rambleed}, but these affect discrete binary states and are comparatively better tolerated, unlike the continuous nature of quantum states.

\subsubsection{Crosstalk induced fault injection}

Crosstalk refers to the unintended interaction between two or more qubits in a quantum computer which can result in errors during computation. These errors can arise from environmental noise or from deliberate attacks by adversaries. Crosstalk can be especially concerning in a MTC environment, where multiple quantum programs may be executed simultaneously on different sets of physical qubits. MTC is economically attractive as it maximizes hardware resource utilization and profitability. Fault-injection attacks can be launched by manipulating the interactions between qubits with the aim of introducing errors into the computation. Such attacks can have substantial socio-economic consequences, for example, a deterministic fault in a weather forecast or an optimal power grid topology calculation could provide an undue financial or political advantage to an adversary.

A crosstalk modeling analysis framework for near-term quantum computers \cite{saki-islped} reveals that crosstalk can be of the same order as gate error, which is considered a dominant error in NISQ devices. Adversarial fault injection using crosstalk in a MTC environment where the victim and adversary share the same quantum hardware has also been proposed. The attack model \cite{saki-islped} assumes that the adversary can run his/her program on the same hardware as one or more victim programs. A conceptual diagram of fault-injection is shown in Fig. \ref{fig:fault-injection}. It is assumed that the adversary, (i) will know the public information e.g., coupling map of the hardware; (ii) may also be aware of the crosstalk values between various qubits by running crosstalk characterization experiments such as idle tomography and simultaneous randomized benchmarking on the qubits before the attack; and (iii) can request to run several copies of small quantum circuits so that he can control the maximum number of remaining qubits after the victim's qubits are allocated.
The attack has been demonstrated using both simulation and experiments using a 3-qubit Grover search as the victim program and repeated CNOT drive as the adversary program. The results, collected from \emph{ibmq\_5\_yorktown (ibmqx2)} a Canary processor, (Fig. \ref{fig:fault-injection}) show that the correct output probability of the Grover search falls drastically (i.e., becomes indistinguishable) after a certain number of CNOTs in the adversary circuit.

The quantum nature of crosstalk means that operations on one qubit can affect the quantum states of nearby qubits in ways that have no classical analogue, such as altering phase relationships or creating unwanted entanglement. This interference is far more impactful in quantum systems than in classical ones, as even slight perturbations can collapse or significantly alter quantum states. Crosstalk stems from qubits' sensitivity to noise and excited state dissipation. Efforts to reduce circuit depth by packing gates for parallel execution often worsen crosstalk effects. The scarcity of high-fidelity qubits further complicates mitigation strategies. Disabling affected qubits as a solution may incur significant performance penalties due to the exponential scaling of quantum computational power, making crosstalk a critical challenge in quantum computing.

In classical systems, crosstalk between interconnects has been studied \cite{shepard1996noise}, but it typically only affects signal integrity rather than fundamentally altering the computation as in quantum systems. 
While classical crosstalk is generally seen as a nuisance to be minimized using well-established techniques like shielding, quantum crosstalk presents a potential attack vector that could be exploited to manipulate quantum computations in subtle ways.

\subsubsection{SWAP induced fault injection}

It has been shown that attacker can exploit the connectivity constraints and scheduling policies in shared quantum hardware to force the insertion of additional SWAP operations into a victim's quantum circuit during compilation \cite{upadhyay2024stealthy}. A conceptual diagram of SWAP-injection is shown in Fig. \ref{fig:swap-injection}. The attacker is assumed to have the knowledge of the quantum hardware's connectivity graph and access to submit jobs to the shared quantum computing environment. The attack involves strategically occupying specific qubits by submitting adversarial programs, exploiting scheduling policies to limit the victim's access to well-connected qubits. This forces the victim's circuit to use a suboptimal qubit mapping, resulting in additional SWAP operations being inserted during compilation. A maximum increase of approximately 55\% and a median increase of about 25\% in SWAP overhead has been reported for victim circuits under attack leading to fidelity degradation and in worse cases fault injection.

SWAP-induced fault injection exploits unique characteristics of quantum computing systems that have no direct classical analogue. Unlike classical bits which are uniformly addressable and accessible, qubits in current quantum hardware have limited connectivity, necessitating SWAP operations to entangle distant qubits. The sensitivity of quantum states to these additional operations is far greater than in classical systems, where data movement from one memory location to another only degrades performance but not the computation fidelity.

\subsubsection{Shuttle induced fault injection}

An attack on MTC systems in TI computing \cite{saki2021shuttle} exploits a new vulnerability in terms of shuttle operations between traps. An overview of the attack model is provided in Fig. \ref{fig:TI}b-d. An example attack scenario involves two users submitting their quantum programs to a quantum cloud, with User 1's program having 2 qubits and User 2's program having 6 qubits. In order to maximize resource utilization, the cloud schedules multiple programs on the same hardware which is commonly practiced in commercial quantum clouds such as Rigetti's Quantum Cloud Service. However,  adversary's program may span over multiple traps and share a trap with qubits from the victim's program. For example, adversary and victim qubits share Trap--$1$ (T$1$). The adversary can design their program to require computation between ions from different traps which will require frequent shuttles between traps. This repeated shuttling can add energy to an ion and increase the ion-chain's energy, resulting in degraded reliability of computation (i.e., reduced gate fidelity). As the victim qubits share a chain with the adversary qubits, they also suffer from this shuttle-induced degradation in gate fidelity.

It has been noted that while the premise of the attack seems simple, there are architectural policies in place to curb shuttling, making the naive attack challenging. Nevertheless, the adversary can design a program that can trick these policies and enforce repeated shuttles, and can be launched in either a \emph{white-box} setup where the attacker has prior knowledge of the policies, or in a \emph{black-box} setup where no prior information is known.

\begin{figure}
    \centering
    \includegraphics[width=3.5in]{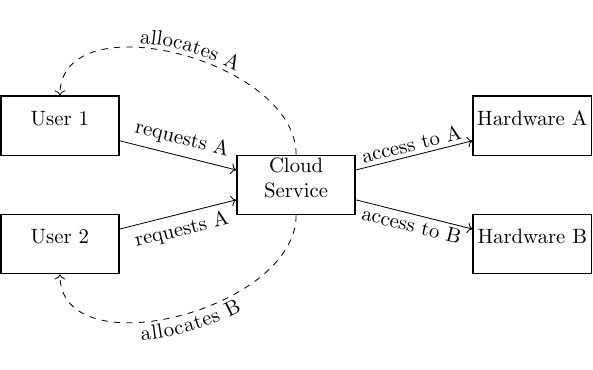}
    \caption{A conceptual attack scenario in which both users request hardware A (which offers superior quality), but User 2 is instead granted access to hardware B.~\cite{puf}.}
    \label{fig:attack_model}
\end{figure}

The attack leverages the fact that excessive shuttling can increase the energy of the entire ion chain, affecting not just the shuttled ion but also neighboring ions. In classical systems, frequent data movement can create congestion and degrade the performance in shared computing environment without affecting the computation outcome unlike quantum systems. The collective effect of shuttling on qubit fidelity is a distinctly quantum phenomenon, arising from the delicate nature of quantum superposition states and the ion-ion interactions.

\subsection{Program rerouting to lower quality hardware}

Quantum circuits are sent to quantum hardware via a cloud-based provider which allocates the hardware for the circuit. Here, the user has no visibility on the hardware that is being allocated. An attack model has been proposed \cite{puf} where the user is allocated an inferior quantum hardware instead of the desired one. Moreover, even if the desired quantum hardware is allocated, the scheduling policy for the queue of quantum circuits is another aspect that should be taken into consideration. The queue of quantum programs on the cloud side is usually long, with the main goal of maximizing throughput for cost reduction and better scientific exploration. The scheduling policy of the hardware is usually provided by the vendor for program allocation to hardware. In Fig. \ref{fig:attack_model}, two users \textbf{U1} and \textbf{U2} request for hardware \textbf{A} which is better compared to hardware \textbf{B} in terms of characteristics like error rates and fidelity. First U1 sends the request to the cloud service, and cloud service allocates hardware A to U1. However, when U2 requests for hardware B, the cloud service could make U2 either wait or allocate hardware B, which is of inferior quality. If the latter happens, U2 will suffer from incorrect results due to inferior hardware and may also end up paying more. In a variant of the scheduler attack \cite{samah-iccad}, the scheduler allocates the requested device however, a rogue employee in the quantum computing company can alter the reported error-rate data so that an inferior qubit (with a larger error rate) is reported as a superior qubit (with a smaller error rate). When a user requests or a compiler allocates physical qubits, they (user or compiler) can unknowingly select inferior quality qubits to run the circuit. Therefore, the user circuit will experience heightened errors and sub-optimal output. It has been stated that the success of the attack depends on the variation of the error rates of the underlying hardware and the failure model applied to determine the mapping policy. An attack can be triggered if there is a significant variation in the qubit error rates.

Unlike classical computers, where computational errors are typically low (with DRAM error rates reported at $10^{-10}$ to $10^{-8}$ per hour per bit \cite{dixit2011impact}), quantum computers are inherently probabilistic and highly sensitive to noise and decoherence, with current error rates around $10^{-3}$ to $10^{-2}$ per gate operation. While classical cloud computing can have security implications, such as potential side-channel attacks when virtual machines share physical servers \cite{ristenpart2009hey}, the impact of resource allocation in quantum systems is far more severe. In quantum computing, the specific qubit allocation can dramatically affect the accuracy and even the possibility of successful computation, not just potentially leak information. This makes the quality of quantum hardware, particularly qubit error rates and fidelity, crucial to computation success. Current NISQ devices can have significant variations in qubit quality within the same chip or between different chips. Furthermore, the complexity of quantum hardware management necessitates opaque cloud services, where users cannot directly verify the quality of allocated resources. This scenario is distinct from classical cloud computing, where performance metrics are more standardized and verifiable, highlighting the unique challenges in quantum resource allocation and verification.


\begin{figure}
    \centering
    \includegraphics[width=3.5in]{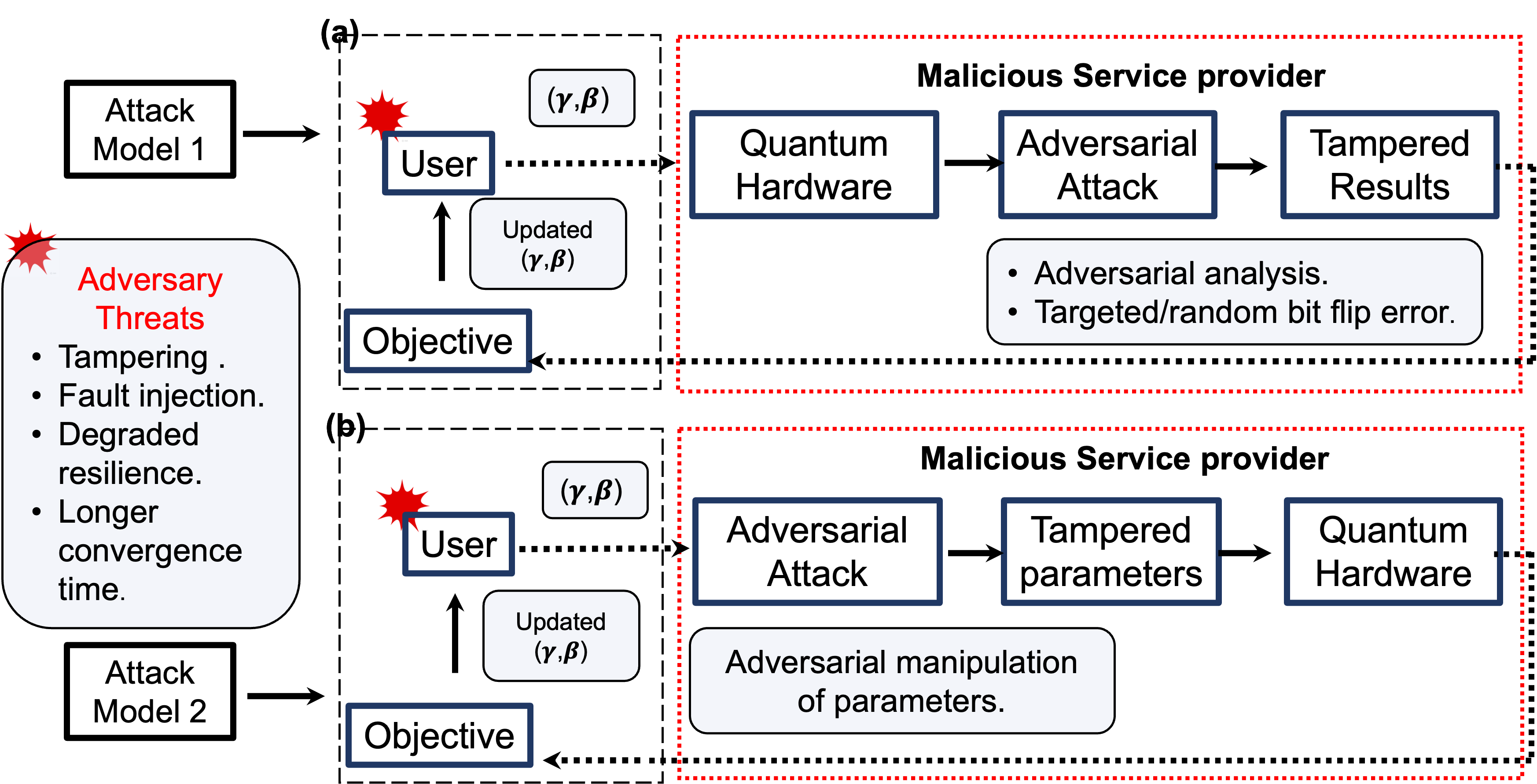}
    \caption{Adversarial tampering attack models~\cite{upadhyay2022robust}. a) Attack model 1 considers an adversary who tampers with the outcome after an iterative program has been run on the hardware. b) Attack model 2 considers an adversary who tampers with the parameters before running the program on a quantum computer.}
    \label{fig:hasp}
\end{figure}

\begin{figure}
    \centering
    \includegraphics[width=3.5in]{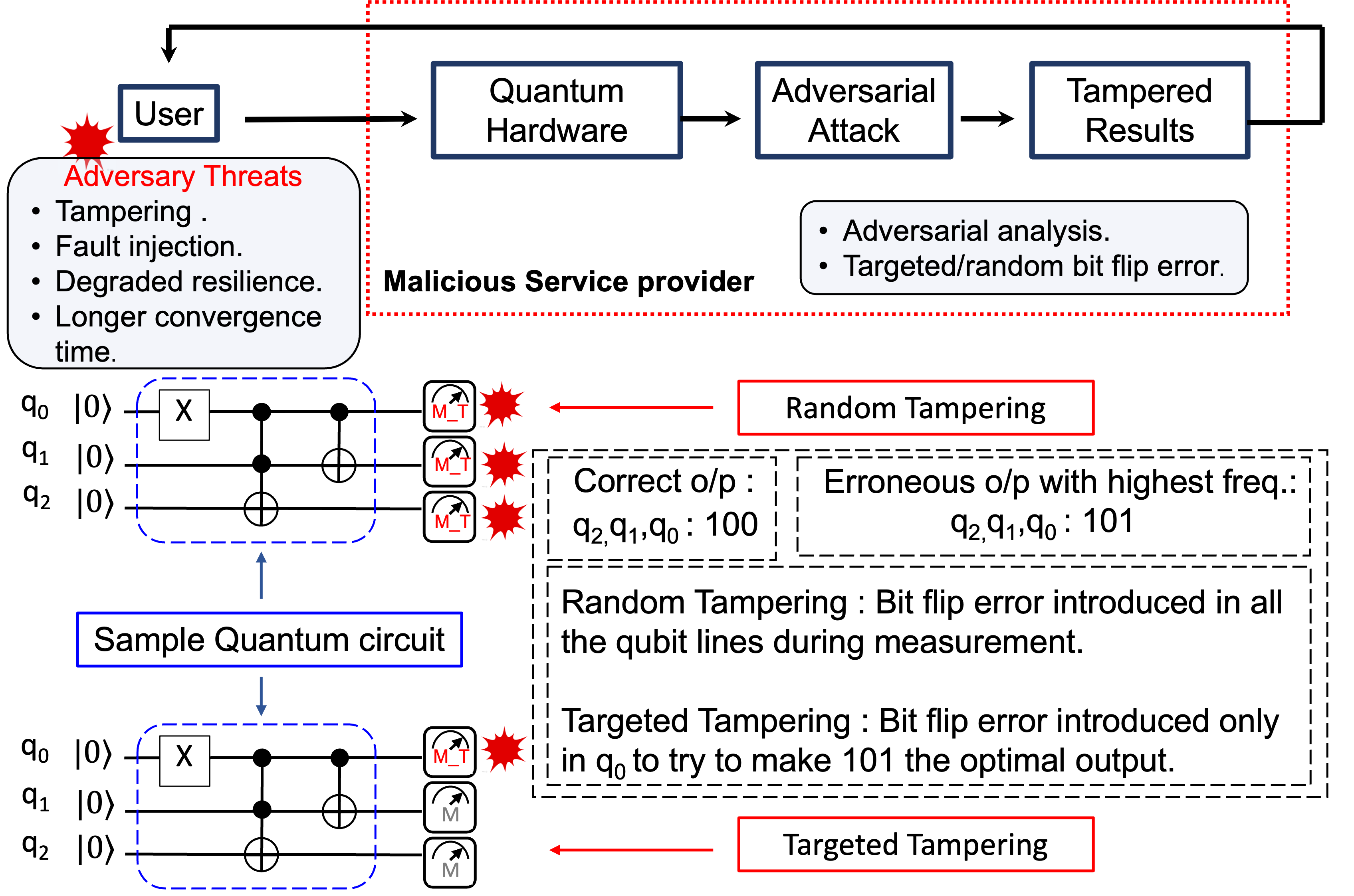}
    \caption{Proposed attack model where the attacker introduces targeted or random tampering, resulting in the users receiving a less-than-optimal solution~\cite{upadhyay2023trustworthy}. Adversarial random tampering and targeted tampering by introducing a bit flip error during measurement.}
    \label{fig:snp}
\end{figure}

\subsection{Adversarial input manipulation}

Similar to the attack models on classical ML algorithms, attacks on QML can be categorized from three dimensions- (i) timing i.e., if the attack take place during training or inference; (ii) information i.e., the type of information that is available to the attacker e.g., knowledge of the internals of the QML model/algorithm (white-box attacks) or access to the inputs/outputs of the QML model only (black-box attacks); and (iii) objectives of the attacker e.g., force misclassification for certain inputs (targeted attacks) or affect the overall reliability of a model (non-targeted attacks).
A non-targeted/reliability attack where an attacker induces noise to a victim's quantum classifier (during inference) through crosstalk in a MTC environment \cite{saki-islped} increased the misclassification rate of the classifier significantly. Ways to generate adversarial samples for a QML image classifier (noisy inputs that are misclassified by the classifier) in both white-box and black-box setup \cite{lu2020quantum} employed additional noise which acted as a unitary to modify the input state to the classifier. However, using such adversarial samples to perform actual attacks is an open research question.

As the popularity of quantum computing grows, more companies, including unreliable vendors, are expected to offer hardware-as-a-service options. Due to the high cost of quantum computing time and the lengthy waiting list for access, users may opt for less expensive but less trustworthy hardware. An attack model where such less-trusted vendors may manipulate the results or parameters of quantum circuits, resulting in subpar solutions for the user or increased costs through a higher number of iterations, has been proposed in \cite{upadhyay2022robust} (Fig.\ref{fig:hasp}). The adversarial tampering of input parameters and measurement outcomes on the Quantum Approximate Optimization Algorithm (QAOA), a hybrid quantum-classical algorithm, has been modeled and simulated. The results show a maximum degradation of performance by approximately 40\%. To achieve comparable performance with minimal tampering, the user incurs a minimum cost of 20X higher iteration.

In hybrid algorithms like QAOA, the interplay between quantum and classical components creates unique attack surfaces. The manipulation of quantum circuit parameters or measurement outcomes exploits the probabilistic nature of quantum measurements and the sensitivity of quantum algorithms to noise, aspects not present in classical optimization. Adversarial attacks on machine learning models have been extensively studied in classical systems. It has been demonstrated that imperceptible perturbations to input images could cause state-of-the-art neural networks to misclassify with high confidence \cite{szegedy2013intriguing}. In QML, similar attack vectors exist, but with unique quantum characteristics. Moreover, the current scarcity and high cost of quantum resources make QML models particularly vulnerable to attacks that increase computational overhead, a concern less prominent in classical ML.

\begin{figure}
    \centering
    \includegraphics[width=3.5in]{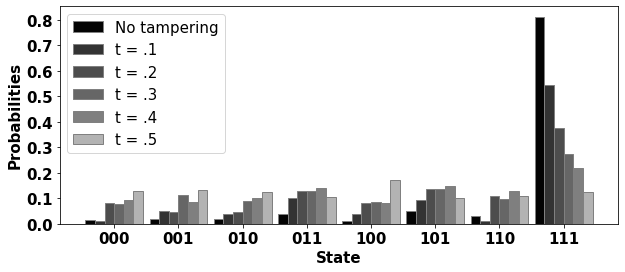}
    \caption{ Sample benchmark (toffoli$_-$n3, correct output = `111') simulated on the fake back-end (Fake$_-$montreal) for 10,000 shots. Changing the tampering coefficient models the extent of adversarial tampering (t). For t =0.5 erroneous state `100' becomes the most occurring output ~\cite{}.}
    \label{fig:snp2}
\end{figure}

\subsection{Adversarial output manipulation}

An attack model in which less-trusted service providers can pose as trustworthy and tamper with the results, leading to sub-optimal solutions being reported to users, has been proposed \cite{upadhyay2023trustworthy}. To demonstrate the impact of this tampering, a simple program has been run on tampered and tamper-free hardware and the probability distributions of basis states have been compared. It has been reported (Fig. \ref{fig:snp2}) that as the tampering coefficient (t) increases, the probability of the correct output '111' decreases while the probability of erroneous outputs increases. In practical scenarios, the user is reliant on the sub-optimal output of the tampered quantum computer since the correct solution to the optimization problem is unknown. Fig\ref{fig:snp} illustrates the proposed attack model.
It has been assumed that the adversary a) has access to the measured results of the program run by the user and b) does not manipulate the quantum circuit to avoid suspicion. However, while measuring the qubit lines, the adversary introduces random bit flip errors or the attack in a more targeted tampering is more strategic in nature, focusing on specific qubit lines.

Unlike classical computing, where bit values are deterministic and easily verifiable \cite{leroy2009formal}, quantum computations rely on probabilistic measurements of superposition states. This inherent uncertainty makes it challenging for users to distinguish between legitimate quantum noise and malicious tampering. Adversaries could potentially introduce errors during the measurement process, exploiting the fact that quantum measurements alter the system state—a stark contrast to classical systems. While classical cloud computing faces its own trust issues \cite{pearson2010privacy}, quantum cloud computing presents more severe challenges due to the scarcity of resources and the difficulty of verification. The complexity of quantum algorithms and current limitations in independently verifying quantum computations make it particularly challenging to detect sub-optimal solutions, especially in optimization problems where the correct answer is unknown. This creates a unique dependency on quantum service providers, as users cannot easily repeat or independently verify computations, unlike in classical systems. Consequently, quantum cloud computing introduces unprecedented trust issues that extend beyond those present in classical cloud environments.

\begin{figure*}
    \centering
    \includegraphics[width=7in]{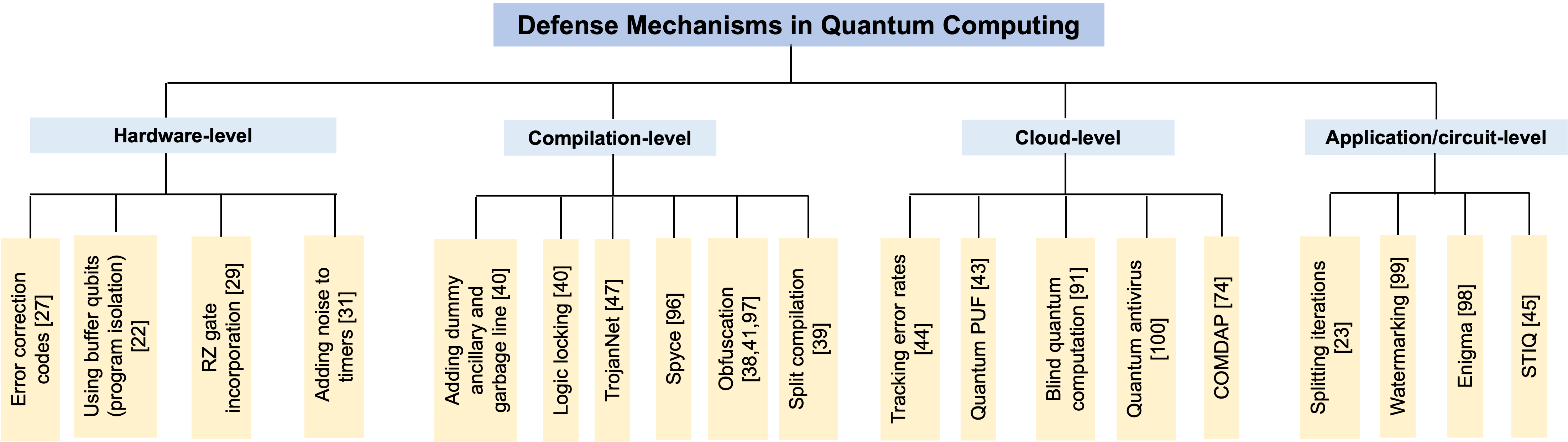}
    \caption{Quantum computing defense model taxonomy. A classification of various defense strategies aimed at securing quantum systems, including hardware, compilation, cloud-based access, and applications (Section VI).}
    \label{fig:defense_taxonomy}
\end{figure*}

\section{Privacy issues in quantum computing}

\subsection{Readout sensing}

The readout error in a quantum computer is state-dependent which means state $\ket{1}$ and $\ket{0}$ experience asymmetric bit-flip probabilities. Besides, the asymmetry extends beyond a single qubit. For example, if 2 qubits are read, then 4 possible states - $\ket{00}$, $\ket{01}$, $\ket{10}$, and $\ket{11}$ - will show asymmetric bit-flip probabilities. Therefore, if the two qubits belong to two different programs each, one adversary and another victim, the adversary can extract information about the state of the victim just by reading his/her qubit. It has been demonstrated that an adversary can exploit readout error and infer another user's output by reading his/her qubit in \cite{saki-tqe-sensing}.
\begin{figure}
    \centering
    \includegraphics[width=3.5in]{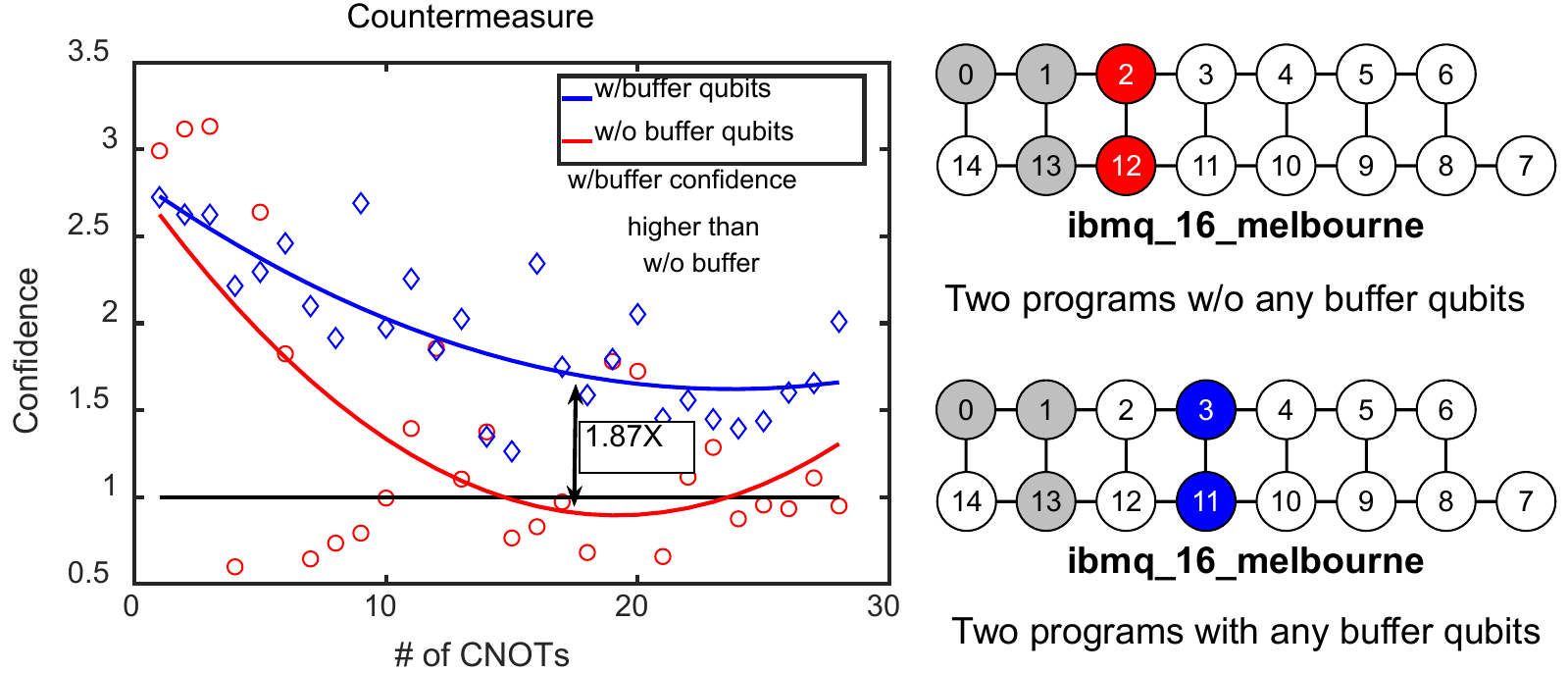}
    \caption{Buffer qubits as a countermeasure against crosstalk induced fault-injection attack.}
    \label{fig:buffer-qubits}
\end{figure}

\begin{figure}
    \centering
    \includegraphics[width=3.5in]{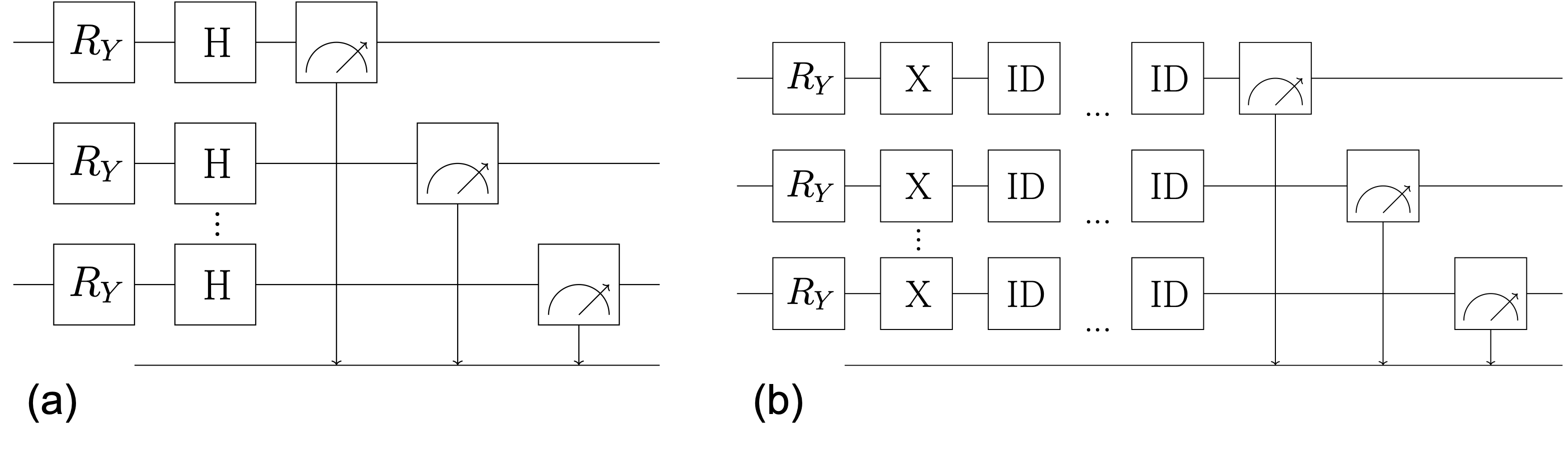}
    \caption{Proposed QuPUFs~\cite{puf}: (a) Hadamard gate-based QuPUF, (b) decoherence-based QuPUF. The tunable rotation has been added for resilience.}
    \label{fig:qupufs}
\end{figure}
Experimental results have shown that the adversary output distribution is glaringly different for victim qubit being $\ket{0}$ and $\ket{1}$. The sensing attack involves two steps, (i) the adversary collects reference signatures from a device by running circuits on both qubits, (ii) the adversary reads only his/her qubit and compares it with the reference signature using a statistical distance (Jensen-Shannon Distance)~\cite{connor2013evaluation}. If the collected signature is statistically closer to reference signature $\ket{1}$ than $\ket{0}$ then it is inferred as $\ket{1}$ and vice-versa. An inferencing accuracy of $96\%$ from experiments has been reported \cite{saki-tqe-sensing}. 

The state leakage attack model has been explored in \cite{xu2024thorough}, where an adversary attempts to extract information about a victim user's quantum circuit execution through imperfect reset operations between consecutive runs on shared qubits. The attack exploits remnants of the victim's quantum state that persist due to imperfect reset operations, allowing potential inference of the victim's computation results or input states through subsequent measurements.

In classical systems, readout errors are typically symmetric and independent of the stored value \cite{schroeder2009dram}. In classical memory technologies like DRAM, error rates can vary but are generally low, with the median error rates of 25,000 to 75,000 FIT (failures in time) per Mbit (about $2.5X10^{-11}$ to $7.5X10^{-11}$ errors per bit per hour)\cite{schroeder2009dram}. Furthermore, these errors are typically not dependent on the stored value (0 or 1). The state-dependent nature of readout errors in quantum computers, where $\ket{0}$ and $\ket{1}$ states experience asymmetric bit-flip probabilities, is a consequence of the quantum nature of qubits and physical implementation of qubit measurement. This asymmetry extends to multi-qubit states, reflecting the complex interactions in quantum systems. The ability to infer the state of one qubit by measuring another exploits quantum entanglement, a phenomenon unique to quantum systems.

\subsection{Reverse engineering and IP theft}

Quantum Circuits are based on reversible logic. Several works on the security of reversible circuits have been conducted \cite{saeed2018ic,cui2018difficulty,limaye2019revisiting,saeed2019reversible}. A potential IP/IC piracy attack on reversible circuits has been identified, along with countermeasures \cite{saeed2018ic}. The synthesis of a target Boolean logic function into a reversible circuit, using techniques such as quantum multiple-valued decision diagrams (QMDD) or binary decision diagrams (BDD), can add ancillary lines on the input side and garbage lines on the output. These added lines offer a level of inherent privacy, as an adversary would need to know their location and values to identify the circuit's functionality. However, the synthesis process can also leave behind signs that can help an adversary locate the ancillary and garbage lines, potentially enabling an attack on the embedded functionality. The risk of end-user piracy and reverse engineering attacks when both the end-user and foundry are untrusted, with access to a netlist and functional chip, has been highlighted in \cite{limaye2019revisiting}.

When utilizing variational algorithms, such as QAOA, to design parametric quantum circuits that are tailored to specific problems, the circuit's topology embodies the problem's structure and can be considered a valuable asset or intellectual property (IP). This IP may not pose a risk for small-scale quantum circuits that can be compiled by trusted vendors, such as IBM and Rigetti. However, with the growth of third-party service providers offering potentially higher performance and the expanding capacity of current NISQ computers, there is an increased likelihood of IP infringement \cite{upadhyay2023obfuscating}. Vulnerabilities in quantum circuit structures used for solving Vehicle Routing Problems (VRPs) with the QAOA have been explored in \cite{chen2024all}. This work focuses on analyzing QAOA quantum circuits that can reveal sensitive information about the problem being solved, such as the locations or connections of military bases or airports in routing optimization scenarios. It has been demonstrated that for random subgraphs with size $\geq 3$, an attacker can reliably recover 95\% of the nodes and match the entire subgraph correctly with probability $\geq 0.975$.

Recent works have investigated vulnerabilities in cloud-based QML systems, particularly focusing on Quantum Neural Networks (QNNs) \cite{kundu2024stiq, ghosh2024quantum}. Both works assume similar threat models involving untrusted cloud quantum computing services with white-box access to the quantum circuits. The risk of model theft in QNNs during both training and inference phases has been addressed in \cite{kundu2024stiq}. The threat model considers an adversary (either the cloud service provider or an insider) with full access to the QNN circuit, including its architecture, trained parameters, and execution details. The potential consequences include replication of the model or offering similar services, thus compromising the intellectual property of the QNN developer.

The vulnerability of transpiled QML circuits to reverse engineering attacks has been studied in \cite{ghosh2024quantum}. This study assumes an untrusted quantum cloud provider with white-box access to the transpiled version of a user-designed, trained QML model during the inference phase. The adversary's goal is to extract the pre-transpiled copy of the QML circuit, which could enable re-transpilation for various hardware and potentially allow further manipulation or refinement of the model. Reverse engineering attempts on various QML models, including multi-qubit classifiers, have been performed. The results demonstrate that under specific conditions these classifiers can be reverse-engineered with a mean error of order $10^{-2}$ in a reasonable time frame.

In quantum circuits, the requirement for reversible logic introduces a novel paradigm where ancillary and garbage lines simultaneously offer privacy and potential attack vectors - a concept alien to classical computing. In classical computing, while circuit design can be proprietary, the structure of the circuit is generic and doesn't typically encode sensitive information about the problem being solved. Classical hardware obfuscation techniques focus on hiding the functionality of the circuit rather than protecting the program itself \cite{baumgarten2010preventing}. Nevertheless, software program can be subjected to various attacks like buffer overflow, code injection and return oriented programming attacks \cite{szekeres2013sok}. For variational algorithms like QAOA, the circuit topology itself becomes sensitive information, encoding problem structures in ways that classical algorithms do not. This creates a unique form of intellectual property in quantum computing, where the mere structure of the circuit can reveal confidential information about the problem being solved, such as critical infrastructure details. Trained QML models are also orders
of magnitude more expensive than classical machine learning (ML) models making them extremely valuable.

\subsection{Quantum circuit identification}

NISQ computers are continuously improving, with larger numbers of qubits and higher fidelity. This improvement makes it possible to execute novel algorithms and generate unique data with valuable intellectual property. However, as quantum computers are usually remote, cloud-based machines, users do not have physical control over them, making them vulnerable to physical attacks from malicious insiders. It has been demonstrated that power-based side-channel attacks could be used to recover information about the control pulses sent to quantum computers \cite{xu2023exploration}. This information can then be used to reverse-engineer the gate-level description of circuits and ultimately the secret algorithms used on the quantum computer. An algorithm for reconstructing a circuit by analyzing per-channel power traces of a quantum computer, assuming the attacker has access to the basis pulse library, has been proposed \cite{xu2023exploration}. The algorithm involves two phases: the search phase and the remove phase. In the search phase, the algorithm converts the power traces into binary format and compares them with the binarized power traces of the basis gates to identify the target gate. In the remove phase, the identified target gate is removed from the power traces, and new power traces are generated for the next iteration of the algorithm. Various types of information that can be recovered from a quantum computer have been discussed, such as identifying user circuits, circuit oracles, circuit ansatz, qubit mapping, quantum processor, and reconstruction from power traces.

Power side-channel vulnerabilities in quantum computer controllers have been explored in \cite{erata2024quantum}. This study investigates usage of power consumption information from the control electronics of quantum computers to reconstruct quantum circuits and extract sensitive information about the algorithms being executed. The focus is on the classical control electronics rather than the quantum processing unit itself. The attack model assumes the attacker has physical access to measure power consumption of the quantum computer's control electronics, without needing direct access to the qubits or quantum processing unit.

Timing-based side-channel vulnerabilities in cloud-based quantum computing services, specifically focusing on IBM's quantum cloud platform \cite{lu2024quantum} show that execution time information can be exploited to infer details about the quantum circuits being run, potentially compromising the confidentiality of quantum algorithms and sensitive data. The attack model assumes that the attacker is an external party with access to the cloud quantum computing service and can submit their own quantum circuits and measure the execution time between submissions. It has been demonstrated that with just 10 measurements, it's possible to identify the specific quantum processor being used. Additionally, different types of victim quantum circuits can be distinguished with a minimum of 1 measurement and a maximum of 18,712 measurements. For popular circuits like Grover's algorithm, the ability to extract the quantum oracle with as few as 500 measurements has been shown, though in some cases up to 20 million measurements may be required.

Side-channel attacks in classical systems, both digital and analog, typically exploit unintended information leakage through physical channels such as power consumption, electromagnetic emissions, or timing variations \cite{lyu2018survey}. However, quantum systems present unique challenges and opportunities for side-channel attacks. Unlike classical systems, where measurement doesn't significantly affect the system state, any observation in a quantum system can collapse the quantum state. The vulnerability to these attacks stems from the current architecture of quantum computers, where classical control systems play a crucial role in manipulating quantum states. This hybrid classical-quantum nature creates new attack surfaces not present in purely classical systems.

\section{Defense mechanisms}

This section presents various defense mechanisms against security threats to the quantum computing ecosystem Fig.\ref{fig:defense_taxonomy}.

\subsection{Preventing crosstalk-induced fault injection via isolation}

\emph{Buffer qubits} has been proposed \cite{saki-islped} to thwart crosstalk-induced fault injection. This countermeasure has been experimentally demonstrated by running parallel circuits on \emph{ibmq\_16\_melbourne} another Canary processor (Fig.~\ref{fig:buffer-qubits}). In scenario--1, two programs were allocated on adjacent qubits: victim program on physical qubits \{Q0, Q1, Q14\} and adversary program on \{Q2, Q12\}. In scenario--2, buffer qubits were introduced between the programs where \{Q2, Q12\} acted as a buffer, and the adversary program was pushed to \{Q3,Q11\}. For scenario--2, higher fidelity (as much as $1.87$x) than scenario--1 has been reported. The use of pulse optimization and avoidance of free qubit evolution as device-level countermeasures has also been suggested.

\begin{figure*} [t] 
 \begin{center}
    \includegraphics[width=1.0\textwidth]{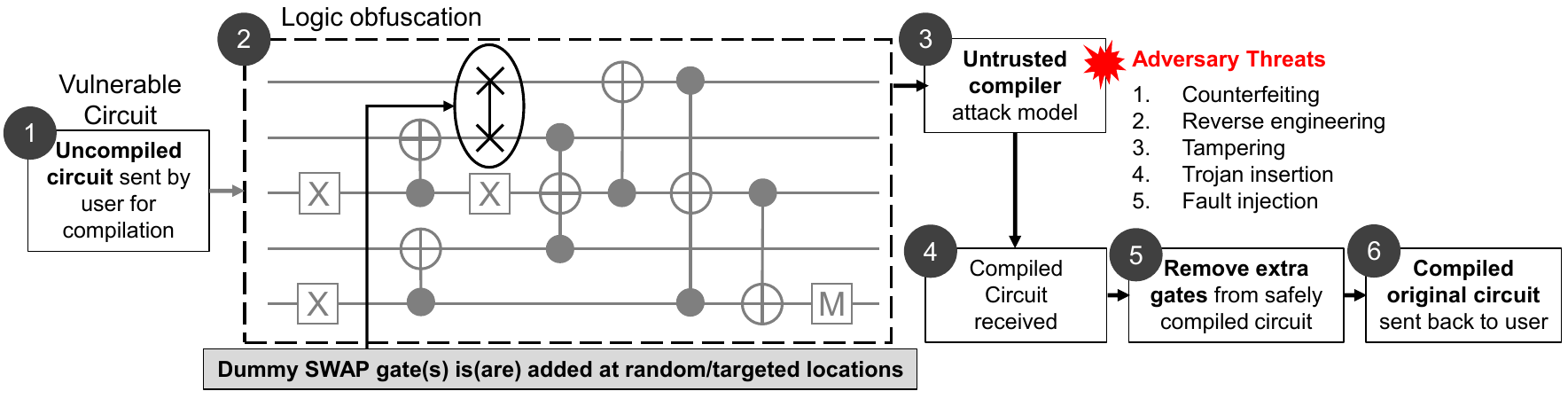}
 \end{center}
 \caption{Attack model proposed in~\cite{obfuscation-aks}. The quantum circuit is sent by the user to the untrusted compiler, where the adversary can steal the IP or reverse engineer the circuit. Logic obfuscation is proposed as countermeasure.} \label{fig:atkmodel}
\end{figure*}

\subsection{Preventing scheduler attack: Tracking changes in error rates}

To detect unexpected changes in error-rates attack, the monitoring of quantum circuit errors using \emph{test points} has been proposed \cite{samah-iccad}. Three different types of tests have been proposed: (i) classical test, (ii) superposition test, and (iii) un-compute test. A user needs to know the expected output to detect any changes in error rates. However, the user does not know output beforehand, otherwise the problem would become trivial. Besides, he/she cannot always resort to simulation since that is computationally expensive. Therefore, the tests are carefully chosen so that the user has knowledge about output. For example, the output of an un-compute test should be the initial state it started with. Two copies of a circuit with test points are run on two isomorphic sub-graphs of the device. The outputs are compared to check if the relative error rates are satisfied. If there is an anomaly, unexpected changes in compile-time information (error rates) are detected.

\subsection{Preventing scheduler attack: QuPUF}

To verify whether the quantum hardware being allocated is the one that is desired or not, the idea of Quantum PUF (QuPUF) has been introduced in \cite{puf}. A QuPUF is a quantum circuit that is sent to the quantum hardware. The parameters of QuPUF and the output given by the hardware act as the challenge-response pair respectively. For each hardware, different challenge-response pairs are accumulated. The assumption here is that each hardware will generate unique challenge-response pairs due to every hardware's unique characteristics like single-qubit error rates, CNOT error rates, decoherence time, and dephasing time. Two  QuPUF models have been proposed namely, Hadamard gate-based QuPUF and decoherence-based QuPUF (Fig. \ref{fig:qupufs}(a) and (b)) as described below.

\subsubsection{Hadamard gate-based QuPUF}

The Hadamard gate-based QuPUF uses the biasing of the probability of the qubits towards either 0 state or 1 state to generate the response. The reason for such biasing could be gate error (usually small for single-qubit gates) or readout error (typically large). At the start, all the qubit states are initialized to a zero state. They are then put in a superposition state using the Hadamard gate, and then the qubits are measured. Ideally, the output should be 50\% probability for both the states. But that won't be the actual case due to the errors and would be biased towards either 1 state or 0 state which would act as a unique device signature.

\subsubsection{Decoherence-based QuPUF}

The decoherence-based QuPUF relies on the decoherence times of the qubits to give unique output. The qubits are initialized to 0 state and then flipped to 1 state using a not gate. The qubits are then allowed to decohere down from 1 state to 0 state by the use of idle gates, which do no operation and simply pass time. In other words, the qubits are excited to a higher state and allowed to decohere down to 0 state. The decoherence of qubits will effectively act as the unique device signature.

\subsection{Securing reversible circuits}

To address the issue of IC/IP piracy \cite{samah-ic-ip-piracy-tvlsi}, two approaches have been proposed. The first approach, which is considered naive, involves adding extra (dummy) ancillary and garbage lines prior to synthesis. After synthesis, additional ancillary and garbage lines are added, but the attack can only identify those added post-synthesis and not the pre-synthesis ones. As a result, the embedded functionality is obfuscated. However, this approach increases the hardware overhead. To minimize the cost, a second approach has been proposed that involves the judicious addition of reversible gates to the circuit, so that after synthesis the ``telltale'' signs are removed keeping the logical functionality intact. To prevent piracy and reverse engineering from the end-use, logic locking has been proposed in \cite{ets-limaye}. In particular, SFLL-HD$^0$, a variant of \emph{stripped functionality logic locking} (SFLL) has been chosen to secure the circuit. The logic locking block consists of 3 sub-blocks: functionally stripped circuit (FSC), restore unit/comparator, and restore signal/XOR. The FSC is formed by either adding or replacing a few logic gates. It inverts the output bit for one protected input pattern (PIP). The comparator/restore unit compares a key and the primary input to generate a restore signal. The key is saved in a tamper-proof memory. Finally, the XOR unit will revert the inverted output depending on the restore signal. The scheme protects against removal and SAT attacks.

\subsection{Blind quantum computation}

Researchers have explored the concept of \emph{blind quantum computation} (BQC) \cite{arrighi2006blind} for preserving the privacy of quantum computation from potentially compromised or malicious servers. Several theoretical protocols \cite{bqc-1, bqc-2, mahadev2020classical} have emerged which allows a client to perform a computation on a server such that the server cannot learn any information about the client's input, output, and computation. Recently classical homomorphic encryption for quantum circuits has been proposed in \cite{mahadev2020classical}. The scheme allows a client to both hide data and performs computation on the hidden data. A review of the BQC protocols is presented in \cite{bqc-review}. Although the theory for BQC is well researched, the physical implementations of such protocols are under-examined \cite{bqc-review}.

\begin{figure*}
    \centering
    \includegraphics[width=1.0\linewidth]{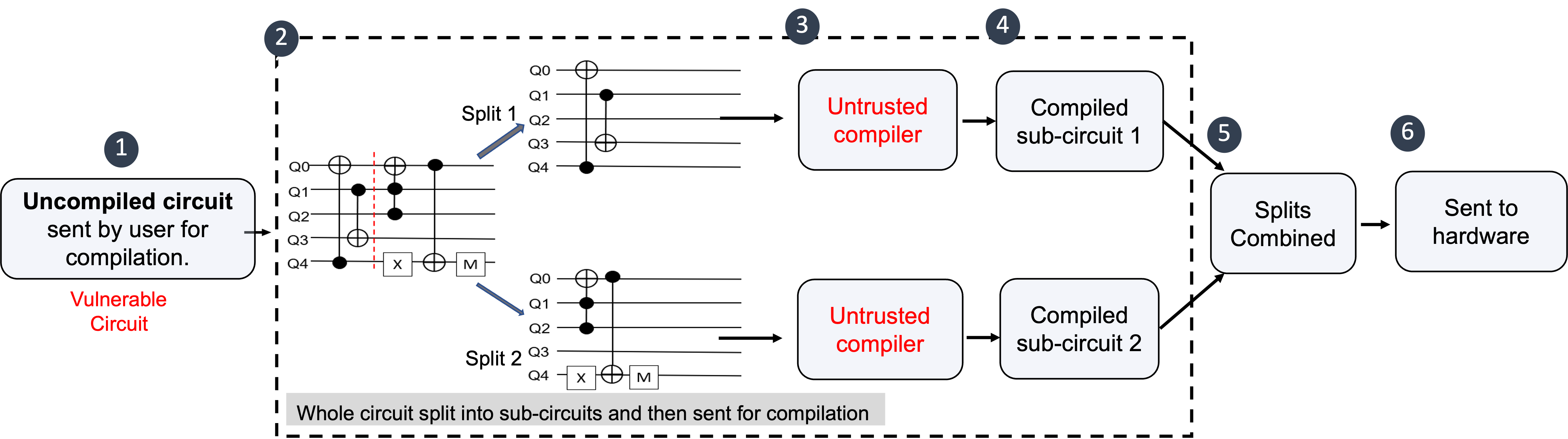}
    \caption{Split compilation technique ~\cite{saki2021split} where the sub-circuits could be sent to the same/different compilers. This method enhances security by protecting intellectual property (IP), as it prevents the exposure of the entire circuit to any single compiler, reducing the risk of unauthorized access or malicious manipulation of sensitive information.}
    \label{fig:split}
\end{figure*}

\subsection{Protection against malicious compiler and hardware providers}

\subsubsection{TrojanNet}

The vulnerability of QAOA circuits to Trojan insertion during compilation by untrusted third parties and TrojanNet, a machine learning approach for detecting and classifying Trojan-inserted circuits, has been developed in \cite{das2023trojannet}. The focus is on QAOA circuits designed for solving the Graph Max-Cut problem, which are popular for various optimization tasks. The assumption is that the user designs a QAOA circuit and employs untrusted or less-trusted third-party compilers to optimize the depth and gate count. The adversary can insert Trojans or extra gates at specific locations to introduce biases or perturbations that disrupt the optimization process, compromising the quality and reliability of the obtained solutions. A Convolutional Neural Network (CNN) model, designed to detect and classify trojan-inserted QAOA circuits, has been introduced. Datasets of Trojan-free and Trojan-inserted compiled QAOA circuits, incorporating variations in Trojan gate types, numbers, insertion locations, and compiler backends, have been generated. The model has been trained and evaluated on the generated datasets, achieving an average accuracy of 98.80\% in differentiating between Trojan-free and Trojan-inserted circuits.

\subsubsection{Splitting the iterations}

An attack model where less-trusted vendors may manipulate the results or parameters of quantum circuits, leading to subpar solutions or increased costs, has been proposed in \cite{upadhyay2022robust}. Adversarial tampering of input parameters and measurement outcomes on an exemplary hybrid quantum classical algorithm, namely, QAOA, has been modeled and simulated. Several countermeasures have been proposed. First, the distribution of computation among multiple available hardware, including a mix of trusted and untrusted devices or multiple untrusted hardware from various vendors, has been suggested. By combining the results from each hardware and iteration, the user can obtain the probability distribution of the solution space. Second, an intelligent, run-adaptive iteration distribution to differentiate between trusted and untrusted hardware and allocate more iterations to the trusted hardware to improve computation quality has been proposed. Third, the re-initialization of the parameters for the hybrid quantum-classical algorithm QAOA after a few initial iterations to counteract any errors introduced by the adversary during the early stages of the computation has been proposed. In their chosen performance metrics, a maximum improvement of $\approx$30\% using their proposed heuristics has been reported.

\subsubsection{Spyce}

Spyce, a novel technique to obfuscate quantum program output and circuit structure on untrusted quantum cloud computing platforms, has been proposed in \cite{patel2023toward}. Unlike previous approaches that focus on encrypting quantum information over networks for securing quantum programs from third-party compilers, Spyce addresses the scenario where the cloud hardware provider itself is untrusted and may have intentional or unintentional snoopers analyzing program outputs. A method that intelligently obfuscates program output and quantum circuit structure of the original quantum program provided by the user/customer has been introduced. Spyce uses a combination of X-gate injections at the end of the circuit for output obfuscation and RX-gate pair injections throughout the circuit for structural obfuscation. The technique involves dividing the circuit into manageable blocks, synthesizing these blocks, and then combining them to form the full obfuscated quantum circuit. It ensures that the number of CX gates (two-qubit gates) remains the same as the original circuit to maintain similar noise characteristics on real quantum hardware. It has been demonstrated that Spyce successfully obfuscates the output probability distribution and dominant state from adversaries while allowing users to recover the original output with a decoding key. The results show that Spyce achieves high structural divergence from the original circuit, maintains comparable output quality on noisy hardware, and incurs minimal compilation time overhead (2x on average).

\subsection{Securing IPs}

\subsubsection{Obfuscation}

As mentioned in Section~\ref{subsec:cloud}, the quantum circuit can be an IP. The addition of \emph{dummy gates} in a circuit to obfuscate the circuit from an untrusted compiler has been proposed \cite{obfuscation-aks} to hide the true functionality of the circuit from the untrusted compiler. The adversary needs to identify and remove the dummy gates from an obfuscated circuit to extract the original circuit. This is a computationally hard problem since any gate can be a potential dummy gate. Any attempt to reuse the circuit without removing the dummy gates will result in corrupted or severely degraded performance.

Fig.~\ref{fig:atkmodel} conceptually shows the idea with a quantum circuit. The original circuit is divided into layers first. Then, inside each layer possible dummy SWAP insertion locations are identified. For example, if a layer has 3 free qubits, there are $\binom{3}{2} = 3$ choices for dummy SWAP gates giving rise to numerous SWAP insertion locations. However, only one dummy SWAP will be inserted in the original circuit and sent to the untrusted compiler. The aim is to insert a dummy SWAP that will cause significant degradation in the output. Exhaustive simulations with a set of test circuits were first run and the impact of dummy SWAP insertion at each possible location was studied. From the study, a heuristic to find out an optimal SWAP insertion location was developed which tracks several features such as the number of control qubits in the path from the SWAP to a measure qubit and calculates a score for the position. On the basis of the score, the optimal SWAP candidate is selected.

The insertion of a small random circuit into the original circuit for obfuscation, which is then sent to the untrusted compiler, has also been proposed \cite{das2023randomized}. Since the circuit function is corrupted, the adversary (i.e., untrusted compiler) may obtain incorrect IP. However, to avoid incorrect output post-compilation, the inverse of the random circuit is concatenated in the compiled circuit to recover the original functionality. The quality of obfuscation is measured using the Total Variation Distance (TVD) metric. The proposed method achieves TVD of up to 1.92 and introduces minimal degradation in fidelity ($\approx 1\% to \approx 3\%$).

A novel protection scheme to safeguard quantum information by storing a secret key separately has been presented in \cite{rasit}. The key controls a set of special added ancilla qubits, called "lock ancilla qubits". To use this protection method, a user determines the desired security complexity level and inserts the required number of lock ancilla qubits. Selected gates are then replaced with controlled versions, where the control is one of the lock ancillas with value $\ket{1}$. Additionally, dummy control gates can be inserted to select active qubit lines, where the control line is one of the lock ancillas with value $\ket{0}$. The order of lock ancilla qubits is mixed randomly and stored separately from the quantum circuit, with their signals applied before running the quantum circuit.

\begin{figure*}
    \centering
    \includegraphics[width=1.0\linewidth]{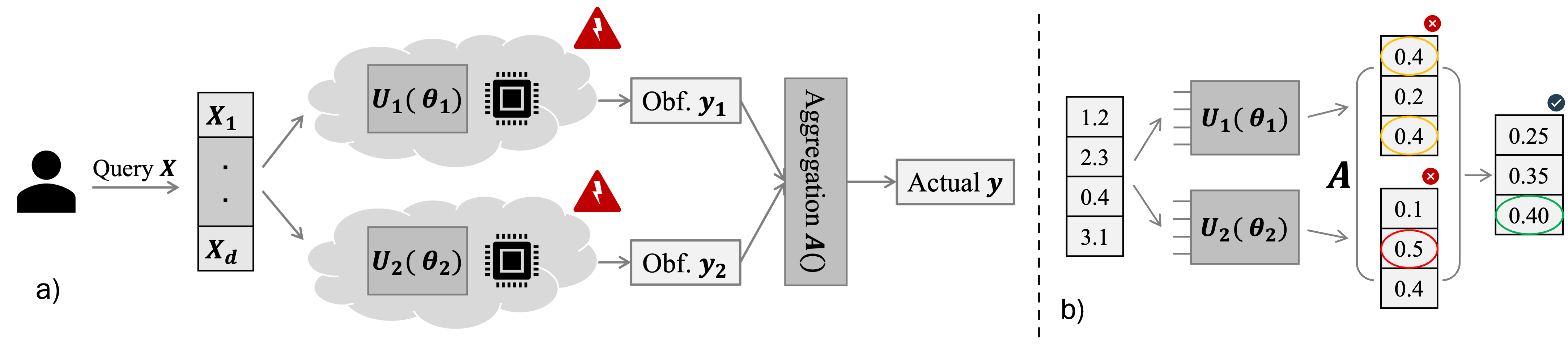}
    \caption{STIQ \cite{kundu2024stiq} inference process: (a) The system processes a user's input query X through two separate quantum neural networks (QNNs) hosted in the cloud. This generates two distinct obfuscated outputs, labeled y1 and y2. STIQ then combines these outputs locally to produce the final correct result y for the user. (b) This section provides an example demonstrating how individual models might generate inaccurate predictions, but when their outputs are aggregated using the STIQ method, the system successfully produces the correct output vector.}
    \label{fig:stiq}
\end{figure*}

\subsubsection{Split Compilation}

A split compilation methodology has been proposed as a solution to secure quantum IPs from untrusted compilers while still utilizing their optimizations \cite{saki2021split}. The methodology involves splitting a quantum circuit into multiple parts, which can either be sent to a single compiler at different times or to multiple compilers. By dividing the circuit in this manner, the adversary only has access to partial information. The sub-circuits will be unscrambled and stitched by the designer post-compilation. Fig~\ref{fig:split} exemplifies the proposed idea.
An analysis of over 152 quantum circuits on three IBM hardware architectures has been conducted in the study. The results showed that the split compilation methodology can secure IPs effectively when multiple compilers are used, or can introduce a factorial time reconstruction complexity while incurring a modest overhead (approximately 3\% to 6\% on average). 

\subsubsection{Enigma}

Enigma, a suite of privacy-preserving schemes specifically designed for the QAOA, has been proposed in \cite{ayanzadeh2023enigma}. Unlike previous Secure Quantum Computing (SQC) techniques that obfuscate quantum circuits, Enigma transforms the input problem of QAOA, such that the resulting circuit and the outcomes are unintelligible to the server. A malicious or compromised server (cloud provider) that has access to the quantum circuit and its execution results is assumed. The server is assumed to be able to distinguish QAOA circuits and potentially reverse-engineer them to infer the problem graph. The aim is to protect all information related to the QAOA problem, including the number of variables, problem coefficients, and the structure of the problem graph. Three variants of Enigma have been introduced. Enigma-I protects the coefficients of QAOA using random phase flipping and fudging of values. Enigma-II protects the nodes of the graph by introducing decoy qubits, which are indistinguishable from primary ones. Enigma-III protects the edge information of the graph by modifying the graph such that each node has an identical number of connections. For all variants of Enigma, it has been demonstrated that the solution for the original problem can still be obtained while keeping the problem structure hidden from the server. Evaluation on IBM quantum devices shows that Enigma's privacy enhancements incur only a minor fidelity reduction of (1\%–13\%).

\subsubsection{Watermarking quantum circuits}

A quantum circuit watermarking approach \cite{samah-watermarking} to detect illegal distribution of quantum circuits/IP infringement embeds a signature in the form of additional gates or hard to remove modified control parameters, while maintaining a high output state fidelity. The approach is based on the unitary gate decomposition of complex operations, and uses a distance metric to guide the error injection for embedding a signature into the quantum circuit. The impact of the proposed watermarking approach on the performance, gate count, and accuracy of the quantum circuit has been analyzed. The integrity of the quantum circuit watermark against removal attacks has also been reported. To evaluate the effectiveness of the proposed approach, QAOA is used as a benchmark to solve a Max-Cut problem.

\subsubsection{Securing quantum neural networks}

Various methods to secure QNNs and QML models against potential threats in cloud-based quantum computing environments have been proposed in recent studies. For example, STIQ (Safeguarding Training and Inferencing of QNNs), an ensemble-based strategy to protect QNNs \cite{kundu2024stiq} involves training two distinct QNNs concurrently, each producing obfuscated outputs that are individually incorrect. These outputs are combined locally using an aggregation function to yield the correct result (Fig. \ref{fig:stiq}). Experiments across various QNNs and datasets have shown that STIQ effectively masks the accuracy and losses of individually hosted models by up to 76\%, with the combined performance matching unobfuscated baseline models. The method introduces a computational overhead of $\leq 2\times$ compared to training a single model. Tests on real 127-qubit IBM\_Sherbrooke hardware have demonstrated up to 60\% obfuscation with combined performance comparable to an unobfuscated model.

Addition of dummy rotation gates with fixed parameters to the QML model as a defense mechanism against reverse engineering has been proposed in \cite{ghosh2024quantum}. Strategies such as increasing the number of layers, adding dummy qubits, or combining both approaches have been suggested. 
Evaluations have demonstrated that adding 2 dummy qubits and 2 layers, increases the reverse engineering overhead by $\sim$1.76 times for a classifier with 2 qubits and 3 layers, with a performance overhead of less than 9\%.

\subsection{Protection for multi-programming environment}

\subsubsection{Quantum antivirus}

The need for new methods to identify malicious circuits before they can be executed on quantum computers has been suggested in \cite{deshpande2022towards}. To defend against fault injection attacks using crosstalk, the implementation of a compile-time technique to scan quantum computer programs for potentially malicious or suspicious code patterns has been proposed. The research (still ongoing) shows that the malicious circuits can take the form of sequences of CNOT gates interleaved with delay gates or a combination of interleaved gates like Pauli X and Y gates. Meanwhile, pure delay gates or I and Z gates do not induce crosstalk errors. Further, pure sequence of CNOT gates is optimized away, so it does not induce crosstalk errors in the victim circuit. The former patterns that can be malicious could be used as the initial set of virus patterns that should be detected by the antivirus during the transpilation process. The extension of the current Qiskit framework for quantum computer programming with antivirus and pattern matching features has been proposed. This can be done by developing an algorithm to count the occurrences of malicious patterns in the quantum circuit and using a database of such patterns to scan user code. For instance, when a low value of K is found, which refers to the number of CNOT gates followed by a delay, it is unlikely that the circuit is malicious. However, if the count is higher than 5 or 10, it may be considered potentially malicious.

\subsubsection{Securing multi-programming environment in super conducting systems}

To protect against adversaries exploiting scheduler policies, a machine learning model for detecting anomalous user behavior in the job queue has been proposed in \cite{upadhyay2024stealthy}. This model utilizes features such as job request frequency, job concurrency, and resource contention patterns to identify potentially malicious users. It has been suggested that schedulers could employ such a model to dynamically adjust priorities or isolate suspicious users, thereby mitigating the risk of attacks.

In another study \cite{upadhyay2024share}, a novel partitioning and allocation method called Community-Based Dynamic Allocation Partitioning (COMDAP) and its secure variant, Secure COMDAP, have been introduced. These methods leverage community detection algorithms to enhance multi-programming capabilities while managing security risks. A Connectivity and Reliability Index (CRI) metric has been introduced to evaluate partition quality and strategies like program padding and smart padding have been implemented to reduce interprogram crosstalk. Secure COMDAP identifies and monitors qubit pairs with high error rates when operating concurrently using Simultaneous Randomized Benchmarking (SRB). By implementing general/smart padding, it places a protective buffer around highly susceptible qubit pairs, preventing their simultaneous operation with other pairs. This approach effectively reduces crosstalk and safeguards against adversarial programs that might exploit it to induce faults. Unlike previously proposed program allocation heuristics that prioritize programs with a high number of CNOT gates, COMDAP aims to distribute resources equitably. This strategy complicates an adversary's attempt to target specific qubits and initiate a SWAP injection attack.

\subsubsection{Securing multi-programming environment in TI systems}

An attack on multi-programming systems in TI computing has been examined in \cite{saki2021shuttle}. To counter this attack, several countermeasures have been proposed:
a) \emph{Random initial mapping:} The creation of malicious programs based on consistent initial mapping has been examined, meaning the same program is allocated in the same fashion every time it is run. To counter this, a solution where the compiler randomly allocates each program at each iteration has been proposed. If a random attack is carried out and the initial mapping changes randomly from one instance to another, the attack program generated at one instance will not work effectively in another instance with a different mapping. The random mapping renders the systematic method useless, as traps will start from unknown states. b) \emph{Dummy pad qubits:} To protect the program, the addition of dummy qubits to pad the unused qubits in a trap, preventing shuttle-induced fidelity degradation, has been suggested. As an example, suppose a user program requires 10 qubits and it is to be executed on a system with a trap capacity of 15 qubits. In this scenario, the user can add 5 dummy qubits to the program, bringing its size to 15 qubits and fully occupying the trap. c) \emph{Capping maximum number of allowed shuttles:} It has been proposed that the cloud provider can implement a maximum limit on the number of shuttles. The cloud provider can examine the required number of shuttles in a program and schedule separate execution if it exceeds the limit. This switch to single-programming mode may result in decreased throughput.

\subsection{Protection against power and timing side-channel attacks}

To defend against power-based side-channel attacks, the incorporation of RZ gate into the original circuits to create a logically equivalent new circuit has been proposed \cite{xu2023exploration}. In contrast to other basis gates that necessitate calibrated pulses, RZ gate can be implemented as a virtual gate via arbitrary wave generators (AWG). The advantage of implementing RZ gate virtually is that it is "perfect" since no actual pulses are required, which means it takes no time to execute. Therefore, RZ gate is undetectable in power-side channels of quantum devices designed to use it virtually. To impede the attacker's ability to reconstruct the original circuit from power traces, the modification of the original circuit by randomly substituting some gates with equivalent sequences that contain RZ gates has been suggested. The attacker encounters difficulty in reconstructing the original circuit because they lack knowledge of the RZ gate's location and rotation angle. This modification improves the circuit's security and protects against power-side channel attacks.

In \cite{lu2024quantum}, it has been shown that execution time information can be exploited to infer details about the quantum circuits being run, potentially compromising the confidentiality of quantum algorithms and sensitive data. Several potential defense strategies have been suggested: 1) Introducing randomness in the compilation process to alter execution times. 2) Implementing a spatial multi-programming system to optimize qubit usage and disrupt timing patterns. This approach divides the quantum processor into regions, running multiple circuits concurrently to obscure individual circuit timings. 3) Modifying quantum circuits slightly to change their timing signatures. This includes adding redundant gates, reordering commutable gates, or using equivalent gate sequences to alter timing profiles without affecting functionality. 4) Adding controlled noise to timers and enforcing privileged access to specific users. This combines introducing noise into timing measurements with restricting high-precision timing data access to authorized users only.

\section{Conclusion and future outlook}
Quantum computing is an emerging paradigm with evolving qubit technologies, hardware architectures, software stack, algorithms, supply chain and usage models. As such, it is important to investigate the assets and IPs embedded in the whole computing stack as well as various vulnerabilities urgently. This is utmost important given that various security/privacy sectors are already exploring the application of quantum computing to solve their problems. The analysis of quantum computing stack through the lens of security has just started with handful of attack models and defense strategies that at best cover only a small fraction of the overall attack surface. Further research is required along the following directions:

\textbf{Supply chain:} The supply chain for quantum computing hardware is not well established compared to the classical hardware counterpart \cite{das2023sok}. Hardware developers use the external vendors and service providers as per their requirements. For example, fabless companies employ third party fabrication facilities to manufacture the quantum chip whereas companies with fabrication house such as, Intel manufacture their own chip. Similarly, dilution refrigerator and calibration services are often procured from third parties. Other hardware components required for quantum hardware include control electronics and cryogenic chips that could be developed in-house or procured from external suppliers. Detailed studies are required to identify the components used in quantum hardware and their typical procurement process. This can inform the potential involvement of untrusted parties/components. 

\textit{IBM's quantum hardware supply chain as a case study:} IBM's quantum computing supply chain involves various components, including hardware, software, services, peripherals, and firmware. IBM sources superconducting materials from suppliers such as, Superconductor Technologies Inc. and American Superconductor Corporation. The design and development of the quantum computing hardware, including qubits/quantum processors (Fig. \ref{fig:stack}.1), is carried out by the IBM Research Center at the IBM Research headquarters and at the Thomas J. Watson Research Center in Yorktown Heights Albany, New York \cite{124}. These materials are processed into wafers at IBM headquarters. The wafers are then fabricated into chips and assembled into the cryostat at IBM's facilities in New York and California. Rigorous testing and quality control processes are conducted at every stage of the supply chain to ensure performance and reliability. The peripherals (Fig. \ref{fig:stack}.3) used in IBM's quantum computing systems include dilution fridges, which are sourced from Bluefors (Helsinki, Finland)\cite{125} and Leiden Cryogenics (Netherlands)\cite{126}. Other peripheral suppliers such as, Agilent Technologies, Keithley Instruments, Keysight Technologies, Pfeiffer Vacuum, and Quantum Design provide necessary equipment for the production and maintenance of the quantum hardware. IBM also partners with other software companies, such as, Zapata Computing, Strangeworks, QxBranch, etc., to provide additional software tools (Fig. \ref{fig:stack}.4) for quantum computing \cite{127}. End-users (Fig. \ref{fig:stack}.5) can access the quantum hardware via cloud-based access. Although IBM's supply chain involves minimal number of third parties, other companies and qubit technologies may involve more number of third parties many of whom could be less reliable or trusted.

\textbf{Qubit technologies:} Current security studies revolve around prominent qubit technologies such as, superconducting qubit and TI qubit. However, other promising qubit technologies and associated toolchains are also being developed concurrently such as, Silicon spin qubit from Intel and nitrogen vacancy and topological qubits developed by multiple entities in parallel. Various qubit technologies may require their customized supply chain, peripherals and software stack. Furthermore, they may possess technology specific vulnerabilities that may enable new mode of attack.   

\textbf{Hardware systems and peripherals:} Quantum chip and peripherals need thorough investigation through the lens of security. In particular, the IP embedded in the hardware implementation needs to be well understood. It is worth noting that multiple architectures exist even for the same qubit technologies. For example, superconducting qubits from IBM and Rigetti may even though functionally look identical (i.e., use Josephson Junction and a capacitor), they may employ different implementation (e.g., microwave frequencies) and peripherals. The native gate sets may also differ. The hardware system and peripherals can be substantially different for another qubit technology.` 

\textbf{Software stack:} Various components of the software stack and the involvement of third party Application Programming Interfaces (APIs), tools and services should be examined for IP and vulnerabilities.   

\textbf{Algorithms:} Quantum algorithms may embed problem specification and other IP/private information that should be thoroughly understood. If possible, various approaches to eliminate, truncate or obfuscate the private information should be developed. 

\section*{Acknowledgements}
This work is supported in parts by NSF (CNS-1722557, CNS-2129675, CCF-2210963, CCF-1718474, OIA-2040667, DGE-1723687, DGE-1821766, and DGE-2113839) and Intel's gift. We also thank the students in LOGICS lab and our numerous collaborators.

\bibliographystyle{IEEEtran}
\bibliography{IEEEabrv,ref}


\begin{IEEEbiography}[{\includegraphics[width=1.05in,height=1.25in,clip]{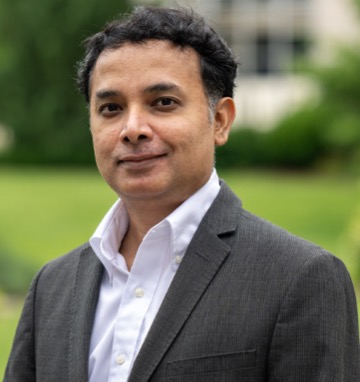}}]
{Dr. Swaroop Ghosh} is a Professor at Pennsylvania State University with research interests in emerging memory technologies, hardware security and quantum computing. He holds a Ph.D. from Purdue University. Before joining academia, he worked as a Senior Research and Development Engineer at Intel Corp. 

Dr. Ghosh has served as an Associate Editor of IEEE Computer Architecture Letters, IEEE Computers, IEEE Transactions On Circuits and Systems I and IEEE Transactions On Computer-Aided Design, and as Senior Editorial Board member of IEEE Journal of Emerging Topics on Circuits and Systems (JETCAS). Dr. Ghosh has also served on technical program committees of over 25 ACM/IEEE conferences, including Microarchitecture (MICRO), IEEE International Symposium on High-Performance Computer Architecture (HPCA), Design Automation Conference (DAC), International Conference on Computer-Aided Design (ICCAD), and Custom Integrated Circuits Conference (CICC). He has received numerous awards for excellence in research, advising and teaching such as Intel Departmental Recognition Award, Intel Divisional Recognition Award, Defense Advanced Research Projects Agency (DARPA) Young Faculty Award, Association for Computing Machinery (ACM) Special Interest Group on Design Automation (SIGDA), Joel and Ruth Spira Award for Excellence in Teaching, Monkowski CAREER Development Professorship, PSEAS Outstanding Research Award, PSEAS Outstanding Advising Award, IEEE-CS TCVLSI Mid-Career Award and NAGS Geoffrey Marshall Mentoring Award. He has also received a Best Paper Awards in ACM International Conference of the Great Lakes Symposium on VLSI (GLSVLSI), 2024 and American Society of Engineering Education Annual Conference, 2020.  Dr. Ghosh is a Fellow of IEEE and AAIA and a Distinguished Speaker of ACM.
\end{IEEEbiography}

\begin{IEEEbiography}[{\includegraphics[width=1.05in,height=1.25in,clip]{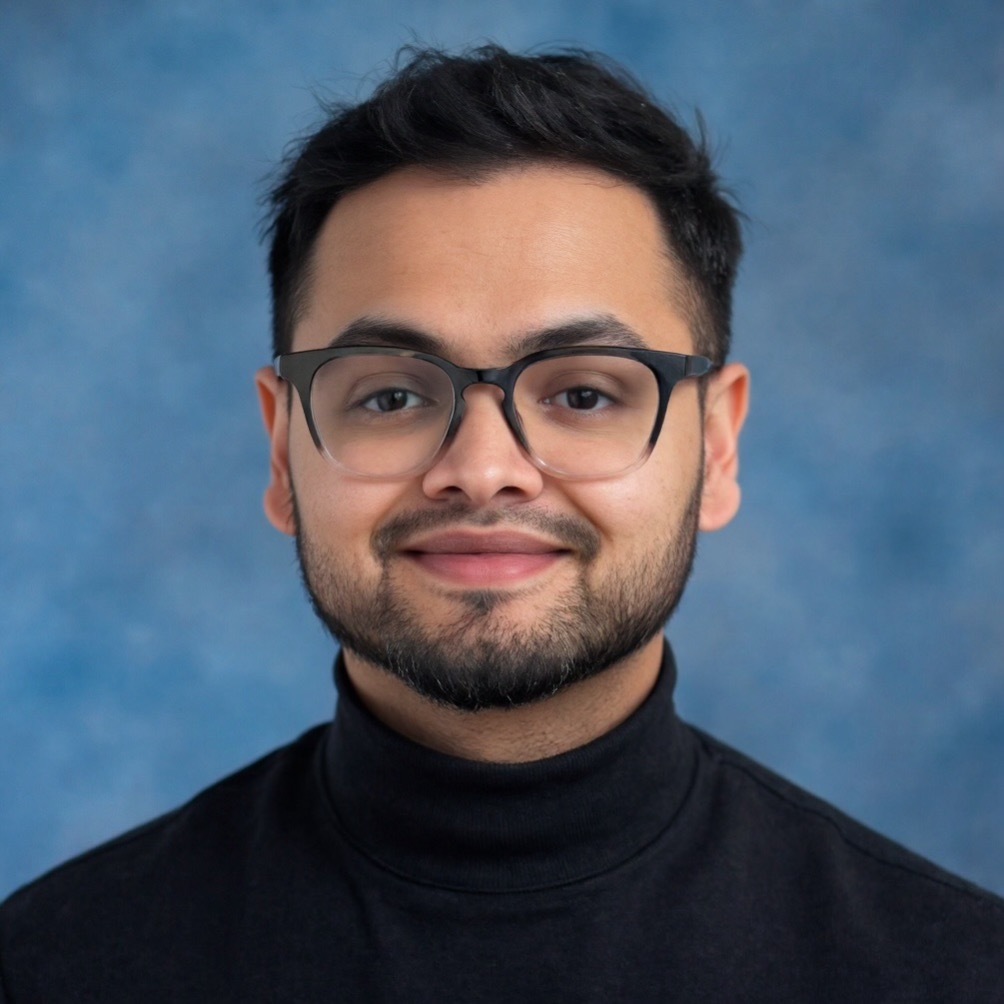}}]{Suryansh Upadhyay} is a Ph.D. candidate in Electrical Engineering at The Pennsylvania State University. His research focuses on quantum computing security, adversarial robustness in quantum machine learning, and optimizing multi-tenant quantum computing frameworks. He has contributed to the field through high-impact publications, security frameworks, and quantum-enhanced learning methodologies. He has served as a reviewer for top-tier journals and conferences and is a Technical Program Committee member for QCE. He has also been awarded the Melvin P. Bloom Memorial Graduate Fellowship for academic and research excellence. He is also a recipient of Dr. Nirmal K. Bose Dissertation Excellence Award from the Department of Electrical Engineering, Penn State.

\end{IEEEbiography}

\begin{IEEEbiography}[{\includegraphics[width=1.05in,height=1.25in,clip]{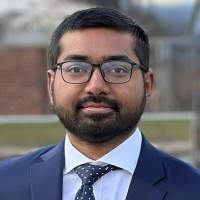}}]{Abdullah Ash Saki} received the B.Sc. degree from the Bangladesh University of Engineering and Technology (BUET), in 2014, and the M.S. and Ph.D. degrees from The Pennsylvania State University, University Park, PA, USA, in 2020 and 2021, respectively. In his Ph.D., he worked on noise resilient and secure quantum computing. He is currently a quantum algorithm engineer at IBM Quantum, where he works on running quantum algorithms at utility scale for practical problems. Prior to that, he was a quantum research scientist at Zapata Computing working on quantum error mitigation techniques to fight noise in NISQ devices. He was a recipient of Dr. Nirmal K. Bose Dissertation Excellence Award from the Department of Electrical Engineering, Penn State.

\end{IEEEbiography}

\vfill



\end{document}